\documentclass[aps,prd,twocolumn,superscriptaddress,preprintnumbers,nofootinbib]{revtex4-2}
\usepackage{amsmath}
\usepackage{amssymb}
\usepackage{natbib}
\usepackage{graphicx}
\usepackage{epsf}
\usepackage{subfigure}
\usepackage{color}
\usepackage{threeparttable}
\usepackage{dcolumn}
\usepackage{comment}
\usepackage{epsfig}
\usepackage{xspace}
\usepackage{multirow}
\usepackage{makecell}
\usepackage{appendix}
\usepackage[
colorlinks=True,
allcolors=blue,
linktocpage=true,
]{hyperref}
\usepackage{latexsym}
\usepackage{array}
\date{\today}

\newcommand\aap{{A\&A}}

\newcommand\apjl{{\it Ap.\ J.\ Lett.\ }}

\DeclareGraphicsExtensions{.jpg,.pdf,.png,.eps,.ps}

\newcommand*{\TT}{\ensuremath{TT}}
\newcommand*{\EE}{\ensuremath{E\!E}}
\newcommand*{\TE}{\ensuremath{T\!E}}
\newcommand*{\lcdm}{$\Lambda$CDM}
\newcommand*{\planck}{\textit{Planck}}
\newcommand*{\polarbear}{\textsc{polarbear}}

\newcommand*{\Neff}{\ensuremath{N_{\mathrm{eff}}}}
\newcommand*{\Yp}{\ensuremath{Y_{\mathrm{P}}}}
\newcommand*{\mnu}{\ensuremath{\Sigma m_{\mathrm{\nu}}}}
\newcommand*{\meff}{\ensuremath{m^{\rm{eff}}_{\rm{\nu, sterile}}}}
\newcommand*{\curv}{\ensuremath{\Omega_{K}}}
\newcommand*{\Hubble}{\ensuremath{H_0}}
\newcommand*{\kmsmpc}{\ensuremath{\mathrm{km\,s^{-1}\,Mpc^{-1}}}}

\newcommand{\RNum}[1]{\uppercase\expandafter{\romannumeral #1\relax}}

\newcommand*{\sqdeg}{\ensuremath{{\rm deg}^2}}

\defcitealias{dutcher21}{D21}
\defcitealias{riess20}{R20}

\definecolor{amber}{rgb}{1.0, 0.49, 0.0}

\newcommand{\skipt}[1]{}

\begin{document}

\title{Constraints on \lcdm{} Extensions from the SPT-3G 2018 $\EE$ and $\TE$ Power Spectra}

\affiliation{School of Physics, University of Melbourne, Parkville, VIC 3010, Australia}
\affiliation{Department of Physics, University of Chicago, 5640 South Ellis Avenue, Chicago, IL, 60637, USA}
\affiliation{Kavli Institute for Cosmological Physics, University of Chicago, 5640 South Ellis Avenue, Chicago, IL, 60637, USA}
\affiliation{School of Physics and Astronomy, Cardiff University, Cardiff CF24 3YB, United Kingdom}
\affiliation{Kavli Institute for Particle Astrophysics and Cosmology, Stanford University, 452 Lomita Mall, Stanford, CA, 94305, USA}
\affiliation{SLAC National Accelerator Laboratory, 2575 Sand Hill Road, Menlo Park, CA, 94025, USA}
\affiliation{Department of Statistics, University of California, One Shields Avenue, Davis, CA 95616, USA}
\affiliation{Fermi National Accelerator Laboratory, MS209, P.O. Box 500, Batavia, IL, 60510, USA}
\affiliation{Department of Astronomy, University of Illinois Urbana-Champaign, 1002 West Green Street, Urbana, IL, 61801, USA}
\affiliation{Department of Physics, University of California, Berkeley, CA, 94720, USA}
\affiliation{Department of Physics \& Astronomy, University of California, One Shields Avenue, Davis, CA 95616, USA}
\affiliation{High-Energy Physics Division, Argonne National Laboratory, 9700 South Cass Avenue., Argonne, IL, 60439, USA}
\affiliation{California Institute of Technology, 1200 East California Boulevard., Pasadena, CA, 91125, USA}
\affiliation{Institut d'Astrophysique de Paris, UMR 7095, CNRS \& Sorbonne Universit\'{e}, 98 bis boulevard Arago, 75014 Paris, France}
\affiliation{Department of Astronomy and Astrophysics, University of Chicago, 5640 South Ellis Avenue, Chicago, IL, 60637, USA}
\affiliation{Department of Physics, Stanford University, 382 Via Pueblo Mall, Stanford, CA, 94305, USA}
\affiliation{Enrico Fermi Institute, University of Chicago, 5640 South Ellis Avenue, Chicago, IL, 60637, USA}
\affiliation{University of Chicago, 5640 South Ellis Avenue, Chicago, IL, 60637, USA}
\affiliation{Department of Physics and McGill Space Institute, McGill University, 3600 Rue University, Montreal, Quebec H3A 2T8, Canada}
\affiliation{High Energy Accelerator Research Organization (KEK), Tsukuba, Ibaraki 305-0801, Japan}
\affiliation{NIST Quantum Devices Group, 325 Broadway Mailcode 817.03, Boulder, CO, 80305, USA}
\affiliation{Materials Sciences Division, Argonne National Laboratory, 9700 South Cass Avenue, Argonne, IL, 60439, USA}
\affiliation{Canadian Institute for Advanced Research, CIFAR Program in Gravity and the Extreme Universe, Toronto, ON, M5G 1Z8, Canada}
\affiliation{CASA, Department of Astrophysical and Planetary Sciences, University of Colorado, Boulder, CO, 80309, USA }
\affiliation{Department of Physics, University of Illinois Urbana-Champaign, 1110 West Green Street, Urbana, IL, 61801, USA}
\affiliation{Department of Physics and Astronomy, University of California, Los Angeles, CA, 90095, USA}
\affiliation{Department of Physics, Case Western Reserve University, Cleveland, OH, 44106, USA}
\affiliation{CSIRO Astronomy and Space Science, PO Box 1130, Bentley WA 6102, Australia}
\affiliation{Department of Physics, University of Colorado, Boulder, CO, 80309, USA}
\affiliation{Department of Physics \& Astronomy, University of Pennsylvania, 209 S. 33rd Street, Philadelphia, PA 19064, USA}
\affiliation{Physics Division, Lawrence Berkeley National Laboratory, Berkeley, CA, 94720, USA}
\affiliation{Department of Physics and Astronomy, Michigan State University, East Lansing, MI 48824, USA}
\affiliation{Three-Speed Logic, Inc., Victoria, B.C., V8S 3Z5, Canada}
\affiliation{Harvard-Smithsonian Center for Astrophysics, 60 Garden Street, Cambridge, MA, 02138, USA}
\affiliation{Dunlap Institute for Astronomy \& Astrophysics, University of Toronto, 50 St. George Street, Toronto, ON, M5S 3H4, Canada}
\affiliation{David A. Dunlap Department of Astronomy \& Astrophysics, University of Toronto, 50 St. George Street, Toronto, ON, M5S 3H4, Canada}
\author[0000-0001-6899-1873]{L.~Balkenhol}
\email[Corresponding author: ]{lbalkenhol@student.unimelb.edu.au}
\affiliation{School of Physics, University of Melbourne, Parkville, VIC 3010, Australia}
\author{D.~Dutcher}
\affiliation{Department of Physics, University of Chicago, 5640 South Ellis Avenue, Chicago, IL, 60637, USA}
\affiliation{Kavli Institute for Cosmological Physics, University of Chicago, 5640 South Ellis Avenue, Chicago, IL, 60637, USA}
\author{P.~A.~R.~Ade}
\affiliation{School of Physics and Astronomy, Cardiff University, Cardiff CF24 3YB, United Kingdom}
\author{Z.~Ahmed}
\affiliation{Kavli Institute for Particle Astrophysics and Cosmology, Stanford University, 452 Lomita Mall, Stanford, CA, 94305, USA}
\affiliation{SLAC National Accelerator Laboratory, 2575 Sand Hill Road, Menlo Park, CA, 94025, USA}
\author{E.~Anderes}
\affiliation{Department of Statistics, University of California, One Shields Avenue, Davis, CA 95616, USA}
\author{A.~J.~Anderson}
\affiliation{Fermi National Accelerator Laboratory, MS209, P.O. Box 500, Batavia, IL, 60510, USA}
\affiliation{Kavli Institute for Cosmological Physics, University of Chicago, 5640 South Ellis Avenue, Chicago, IL, 60637, USA}
\author[0000-0002-0517-9842]{M.~Archipley}
\affiliation{Department of Astronomy, University of Illinois Urbana-Champaign, 1002 West Green Street, Urbana, IL, 61801, USA}
\author{J.~S.~Avva}
\affiliation{Department of Physics, University of California, Berkeley, CA, 94720, USA}
\author{K.~Aylor}
\affiliation{Department of Physics \& Astronomy, University of California, One Shields Avenue, Davis, CA 95616, USA}
\author{P.~S.~Barry}
\affiliation{High-Energy Physics Division, Argonne National Laboratory, 9700 South Cass Avenue., Argonne, IL, 60439, USA}
\affiliation{Kavli Institute for Cosmological Physics, University of Chicago, 5640 South Ellis Avenue, Chicago, IL, 60637, USA}
% ...
\collaboration{SPT-3G Collaboration}
\noaffiliation

\begin{abstract}

We present constraints on extensions to the \lcdm{} cosmological model from measurements of the $E$-mode polarization auto-power spectrum and the temperature-$E$-mode cross-power spectrum of the cosmic microwave background (CMB) made using 2018 SPT-3G data.
The extensions considered vary the primordial helium abundance, the effective number of relativistic degrees of freedom, the sum of neutrino masses, the relativistic energy density and mass of a sterile neutrino, and the mean spatial curvature.
We do not find clear evidence for any of these extensions, from either the SPT-3G 2018 dataset alone or in combination with baryon acoustic oscillation and \planck{} data.
None of these model extensions significantly relax the tension between Hubble-constant, $\Hubble$, constraints from the CMB and from distance-ladder measurements using Cepheids and supernovae.
The addition of the SPT-3G 2018 data to \planck{} reduces the square-root of the determinants of the parameter covariance matrices by factors of $1.3 - 2.0$ across these models, signaling a substantial reduction in the allowed parameter volume. 
We also explore CMB-based constraints on $\Hubble$ from combined SPT, \planck{}, and ACT DR4 datasets.
While individual experiments see some indications of different $\Hubble$ values between the $\TT$, $\TE$, and $\EE$ spectra, the combined \Hubble{} constraints are consistent between the three spectra.
For the full combined datasets, we report $\Hubble = 67.49 \pm 0.53\,\mathrm{km\,s^{-1}\,Mpc^{-1}}$, which is the tightest constraint on $\Hubble$ from CMB power spectra to date and in $4.1\,\sigma$ tension with the most precise distance-ladder-based measurement of $\Hubble$.
The SPT-3G survey is planned to continue through at least 2023, with existing maps of combined 2019 and 2020 data already having $\sim3.5\times$ lower noise than the maps used in this analysis.

\end{abstract}

\keywords{cosmic background radiation -- cosmology: observations -- polarization}

\maketitle

\section{Introduction}
\label{sec:intro}

Measurements of the cosmic microwave background (CMB) provide a unique opportunity to learn about the early universe and its evolution over cosmic time. 
A combination of satellite and ground-based observations have provided a sample-variance-limited view of CMB temperature anisotropy down to few-arcminute scales, beyond which foreground signals dominate \citep{planck18-5, aiola20, choi20, reichardt20}. 
The snapshot of conditions in the early universe provided by the CMB has been crucial in establishing the six-parameter \lcdm{} model as the standard model of cosmology.

Despite its achievements, some questions regarding the \lcdm{} model remain open, such as: is the preference for different cosmologies between large and small angular-scale CMB data physical \citep{planck15-11, planck16-51, henning18, aylor17, addison16}?
What is the origin of the tension between high- and low-redshift measurements of the expansion rate, and can simple model extensions reconcile it \citep{planck18-6, riess20}?
The persistence of these and other tensions, as well as unsolved fundamental physics problems, such as the nature of dark matter and dark energy, is a key motivation for further theoretical study of cosmology \cite{knox19} and construction of more sensitive CMB experiments \cite{abazajian16, ade19}.

Measurements of the CMB polarization on intermediate and small angular scales present an excellent opportunity to investigate these questions. The $E$-mode polarization auto-power spectra ($\EE$) and the temperature-$E$-mode cross-power spectra ($\TT$) contain as much information as the temperature power spectrum ($\TT$) \citep{galli14}, with extragalactic foregrounds relatively dimmer at small angular scales \citep{trombetti18, gupta19, datta19}. 
Thus CMB polarization observations can act both as an important consistency check on the stringent constraints derived from temperature data and as a source of additional and complementary information on the \lcdm{} model and its extensions. 
Improving these measurements is one focus of contemporary ground-based CMB experiments. 
Precision measurements out to few-arcminute scales have been carried out recently by the Atacama Cosmology Telescope (ACT) \citep{choi20}, \polarbear\ \citep{polarbear20}, and the South Pole Telescope (SPT) \citep[][hereafter \citetalias{dutcher21}]{henning18, dutcher21}.

\citetalias{dutcher21} presented $\TE$ and $\EE$ power spectrum measurements from the 2018 observing season of the SPT-3G 1500\,\sqdeg{} survey.
From the SPT-3G 2018 bandpowers, \citetalias{dutcher21} inferred an expansion rate of $\Hubble = 68.8 \pm 1.5\,\mathrm{km\,s^{-1}\,Mpc^{-1}}$, under the \lcdm{} model, in line with other contemporary CMB experiments \citep{planck18-6, aiola20} and lower than the distance-ladder measurement of \citet[][hereafter \citetalias{riess20}]{riess20} using Cepheids and supernovae.
In this paper we consider the implications of the \citetalias{dutcher21} $\TE$ and $\EE$ bandpowers for extensions to the \lcdm{} model.
We assess whether these extensions help reconcile the tension between high- and low-redshift probes of the Hubble constant.

Specifically, we utilise the SPT-3G 2018 bandpower measurements to constrain models with a strong impact on the damping tail, by allowing the effective number of neutrino species, $\Neff$, to vary from the standard model prediction and by breaking big-bang nucleosynthesis (BBN) consistency to change the primordial helium abundance, $\Yp$.
We also constrain the sum of neutrino masses, $\mnu$, the effective mass of one additional sterile neutrino, $\meff$, and spatial curvature, $\curv$. 
While the SPT-3G 2018 bandpowers alone can constrain each of these cosmological extensions, we also look at joint constraints when combined with data from the \planck{} satellite and baryon acoustic oscillation (BAO) data.
After presenting the constraints these datasets place on each model, we investigate the results for $\Hubble$ more closely and discuss any relevant degeneracies in the full parameter space.
Motivated by the higher values of $\Hubble$ inferred from the $\EE$ spectra of contemporary CMB experiments \citepalias{dutcher21}, we look at constraints on the expansion rate from combined measurements of the temperature versus polarization spectra across multiple experiments.
Furthermore, we report the tightest constraint on $\Hubble$ from CMB power spectra to date by combining the temperature and polarization spectra from these datasets, and re-evaluate the Hubble tension.

When analyzing the expansion rate constraints, we choose to compare the CMB results to the distance-ladder measurement of \citetalias{riess20} using Cepheids and supernovae, because of the high precision on $\Hubble$.
We note that the distance-ladder data calibrated using the tip of the red giant branch (TRGB) by \citet{freedman19} agrees with contemporary CMB experiments as well as \citetalias{riess20}, although the TRGB and Cepheid approaches lead to significantly different distances to some supernova-host nearby galaxies \citep{freedman19}.
There are also independent, if more uncertain, constraints on $\Hubble$ using time-delay cosmography \citep{wong19,birrer20}. 
However, for simplicity, we restrict the comparisons in this work to the most precise local measurement of \Hubble{} from \citetalias{riess20}.

This paper is structured as follows. 
In \S\ref{sec:prelim} we review the datasets used in this work and the likelihood used to obtain cosmological parameter constraints. 
We report constraints on \lcdm{} extensions and evaluate their inferred expansion rates in \S\ref{sec:cosmo}. 
We scrutinize Hubble constant constraints from temperature and polarization spectra in \S\ref{sec:t_vs_p} before concluding in \S\ref{sec:conclusion}.

\section{Data Sets and Fitting Methodology}
\label{sec:prelim}

\subsection{The SPT-3G 2018 $\EE/\TE$ Dataset}
\label{subsec:cmb_data}

This work explores the cosmological implications of the first power spectrum measurements from the SPT-3G instrument, which were presented by \citetalias{dutcher21}. 
The $E$-mode auto-spectrum and temperature-$E$-mode cross-spectrum bandpowers are based on observations of a 1500\,\sqdeg{} region taken over four months in 2018 at three frequency bands centered on 95, 150, and 220\,GHz, which result in polarized map depths of 29.6, 21.2, and 75\,$\mu$K-arcmin (averaged across $1000<\ell<2000$), respectively.
The $\EE$ and $\TE$ bandpowers span the angular multipole range $300 \le \ell < 3000$.
Despite the truncated 2018 observing season, the SPT-3G 2018 bandpowers improve on previous SPT results across $300\le\ell\le1400$ for $\EE$ and $300\le\ell\le1700$ for $\TE$ \citep{henning18} and are sample-variance dominated at $\ell < 1275$ and $\ell < 1425$ for $\EE$ and $\TE$, respectively. 
The bandpowers provide precise measurements on the angular scales where hints of physics beyond the standard model may hide.

We adopt the likelihood used in \citetalias{dutcher21}, which accounts for the effects of the aberration due to relative motion with respect to the CMB rest frame \citep{jeong14}, super-sample lensing \citep{manzotti14}, polarized foregrounds, uncertainty in the calibration of the bandpowers, and uncertainty in the beam measurements. 
As in \citetalias{dutcher21}, we place priors on many of these terms, which are listed in Table \ref{tab:priors_table}.
We take a closer look at the effect of super-sample lensing in Appendix \S\ref{app:kappa}.
We refer the reader to \citetalias{dutcher21} for a detailed discussion of the likelihood and the robustness of chosen priors.
The SPT-3G 2018 likelihood will be made publicly available on the SPT website\footnote{\url{https://pole.uchicago.edu/public/data/dutcher21}} and the NASA Legacy Archive for Microwave Background Data Analysis.\footnote{\url{https://lambda.gsfc.nasa.gov/product/spt/index.cfm}}

\begin{table}[ht!]
\def\arraystretch{1.5}
\small
\setlength{\tabcolsep}{10pt}
\centering
\begin{tabular}{l D{+}{\,\pm\,}{-1}}
\hline\hline
 Parameter & \multicolumn{1}{c}{\;\;\;\;\;Prior}\\
 \hline
 $\tau$ & 0.0543+0.0073\\
 $10^3\bar{\kappa}$ & 0+0.45\\
 $A^{\EE}_{80}$ & 0.095+0.012\\
 $\alpha_{\EE}$ & -2.42+0.02\\
 $A^{\TE}_{80}$ & 0.184+0.072\\
 $\alpha_{\TE}$ & -2.42+0.02\\
 $D^\mathrm{ps,~95\times95}_{3000}$ &  0.041+0.012\\
 $D^\mathrm{ps,~150\times150}_{3000}$ & 0.0115+0.0034\\
 $D^\mathrm{ps,~220\times220}_{3000}$ & 0.048+0.014\\
 $D^\mathrm{ps,~95\times150}_{3000}$ & 0.0180+0.0054\\
 $D^\mathrm{ps,~95\times220}_{3000}$ & 0.0157+0.0047\\
 $D^\mathrm{ps,~150\times220}_{3000}$ & 0.0190+0.0057\\
 $T_{\rm cal}^{\rm \,95\,GHz}$ & 1+0.0049\\
 $T_{\rm cal}^{\rm \,150\,GHz}$ & 1+0.0050\\
 $T_{\rm cal}^{\rm \,220\,GHz}$ & 1+0.0067\\
 $E_{\rm cal}^{\rm \,95\,GHz}$ & 1+0.0087\\
 $E_{\rm cal}^{\rm \,150\,GHz}$ & 1+0.0081\\
 $E_{\rm cal}^{\rm \,220\,GHz}$ & 1+0.016\\
\hline
\end{tabular}
\caption[
Priors used for the MCMC fit.
]{
The Gaussian priors listed here are used for the SPT-3G parameter constraints. 
The list of parameters with priors includes the optical depth to reionization $\tau$, mean-field lensing convergence $\bar{\kappa}$, the amplitude $A^{XY}_{80}$ (in $\mu$K$^2$) at 150\,GHz and spectral index $\alpha^{XY}_{80}$ of polarized Galactic dust, the $\EE$ power of Poisson-distributed point sources $D_{3000}^{\mathrm{ps,~\nu_i\times\nu_j}}$ (in  $\mu$K$^2$), absolute temperature calibration factor $T_{\rm cal}^{\, \nu_i}$, and absolute polarization calibration factor $E_{\rm cal}^{\, \nu_i}$.
}
\label{tab:priors_table}
\end{table}

\subsection{Other CMB Datasets}
\label{subsec:other_cmb_data}

We place the SPT-3G 2018 dataset in the wider context of contemporary CMB experiments by comparing its cosmological constraints to the ones produced by ACT DR4 and \planck{} \citep{aiola20, planck18-6}.
The recent ACT DR4 bandpowers \citep{aiola20, choi20} are comparable in constraining power to SPT-3G 2018 while observing a different part of the sky.
The $\EE$ bandpowers of the two experiments are of similar precision across the angular multipole range $300 \le \ell \le 2500$, with ACT DR4 being more precise at $\ell > 2500$.
The ACT DR4 $\TE$ bandpowers are more constraining than the SPT-3G 2018 data across the full angular multipole range.
In contrast to the SPT-3G 2018 data, the ACT DR4 analysis also includes temperature anisotropy measurements.
For the \planck{} satellite \citep{planck18-5, planck18-6}, we use the \textsc{base\_plikHM\_TTTEEE\_lowl\_lowE} set of bandpowers, which are cosmic-variance limited on large to intermediate angular scales. 
Because \planck{} covered the entire sky and does not suffer from atmospheric noise, the \planck{} constraints at low angular multipoles are stronger than those from SPT-3G; conversely, because \planck{} has larger beams and a higher white noise level than SPT-3G, the SPT-3G constraints are stronger at higher $\ell$.
Specifically, the SPT-3G 2018 $\TE$ bandpowers are more precise than the \planck{} data at angular multipoles $\ell > 1400$. 
The \planck{} $\EE$ bandpower uncertainties are smaller up to $\ell < 800$, while the SPT-3G 2018 $\EE$ bandpowers yield better constraints at angular multipoles $\ell > 1000$.

In addition to these three main CMB datasets, we also compare the SPT-3G 2018 constraints to the results from SPT-SZ and SPTpol \citep{story13, henning18} when probing the consistency between temperature and polarization data. 
We do not look at joint parameter constraints from all three sets of SPT bandpowers due to the significant sky overlap between the surveys. 

\subsection{BAO Datasets}
\label{subsec:other_data}

Baryon acoustic oscillation (BAO) measurements provide information about the expansion history of the universe at late times, which is particularly useful to break degeneracies in the CMB data for model extensions that affect the late-time dynamics \citep{efstathiou99, stompor99}.
This class of models is of particular interest in the context of the Hubble tension.
We use BAO measurements from the BOSS MGS and 6dFGS surveys, which have mapped the low-redshift universe in great detail \citep{beutler11, ross15, alam17}. 
We also include the BOSS measurements of the Lyman--$\alpha$ forest and quasars at higher redshifts \citep{blomqvist19}.
Together these datasets provide a detailed view of the expansion history of the universe across $0.2 < z < 3.5$. 

\subsection{Fitting Methodology}
\label{subsec:mcmc}

We produce cosmological constraints using the Markov Chain Monte Carlo (MCMC) package \textsc{CosmoMC} \citep{lewis02b}.\footnote{https://cosmologist.info/cosmomc/} 
\textsc{CosmoMC} uses the Boltzmann code \textsc{camb} \citep{lewis00}\footnote{https://camb.info/} to calculate CMB power spectra at each point in parameter space.
We use the following parameters to describe the \lcdm{} model: the density of cold dark matter, $\Omega_c h^2$; the baryon density, $\Omega_b h^2$; the optical depth to reionization, $\tau$; the (approximated) angular scale of the sound horizon at decoupling, $\theta_{MC}$; the amplitude of primordial density perturbations, $A_s$, defined at a pivot scale of $0.05\,\mathrm{Mpc^{-1}}$; and the scalar spectral index, $n_s$. 

\citetalias{dutcher21} presented constraints on the \lcdm{} model from the SPT-3G 2018 dataset individually and jointly with \planck{} and BAO data. 
We expand that analysis by considering one- and two-parameter extensions to the \lcdm{} model, drawn from these five parameters: the effective number of neutrino species, $\Neff$; the primordial fraction of baryonic mass in helium, $\Yp$; the sum of neutrino masses, $\mnu$; the effective mass of sterile neutrinos, $\meff$; and the spatial curvature, parametrized by $\curv$.
The uncertainties reported in this work on these and core \lcdm{} parameters are 68\% confidence levels.

The optical depth to reionization is constrained primarily by the reionization bump at $\ell<10$ in polarization. 
Since these angular scales are not probed by the ground-based CMB experiments in this work, we adopt a \planck{}-based prior of $\tau = 0.0543 \pm 0.007$ \citep{planck18-6} for all chains that do not include \planck{} data. 
Without this $\tau$ prior, the ground-based CMB constraints show the expected degeneracy between $\tau$ and the amplitude of primordial density perturbations.
We point out that \citet{aiola20} use the prior $\tau = 0.065 \pm 0.015$, which is why we report slightly different results for ACT DR4.

When reporting joint constraints from SPT-3G 2018, \planck{}, and ACT DR4, we ignore correlations between different datasets, unless we combine \planck{} and ACT DR4 temperature data, in which case we restrict the angular multipole range of the latter to $\ell > 1800$ as recommended by \citet{aiola20}.
As the SPT-3G 1500 deg$^2$ observation region is a small fraction of the \planck{} survey field, we expect the covariance between the two datasets to be negligible.
The two ground-based surveys, SPT-3G and ACT DR4, observe different parts of the sky.

\section{Cosmological Constraints}
\label{sec:cosmo}

We now present constraints on extensions to \lcdm{}.
We begin by looking at three extensions that test for new light relics or inconsistencies with BBN: varying the effective number of neutrino species, $\Neff$ (\S\ref{subsec:Neff}); varying the primordial helium abundance, $\Yp$ (\S\ref{subsec:Yp}); or varying both parameters (\S\ref{subsec:Neff+Yp}).
We then turn our attention to questions about neutrino mass, and examine constraints on the sum of neutrino masses, $\mnu$ (\S\ref{subsec:mnu}), and an effective sterile neutrino mass, $\meff$ (\S\ref{subsec:meff}). 
Finally, we discuss the implications of the SPT-3G 2018 data for the spatial curvature parameter, $\curv$, in \S\ref{subsec:curv}.
We highlight key results in this section and refer the reader to Appendix \S\ref{app:tables} for tables containing the full cosmological parameter constraints.

\subsection{Effective Number of Neutrino Species, $\boldsymbol{N_{\rm eff}}$}
\label{subsec:Neff}

\begin{figure*}[ht]
  \centering
  \begin{subfigure}{}
    \includegraphics[width=4.33in]{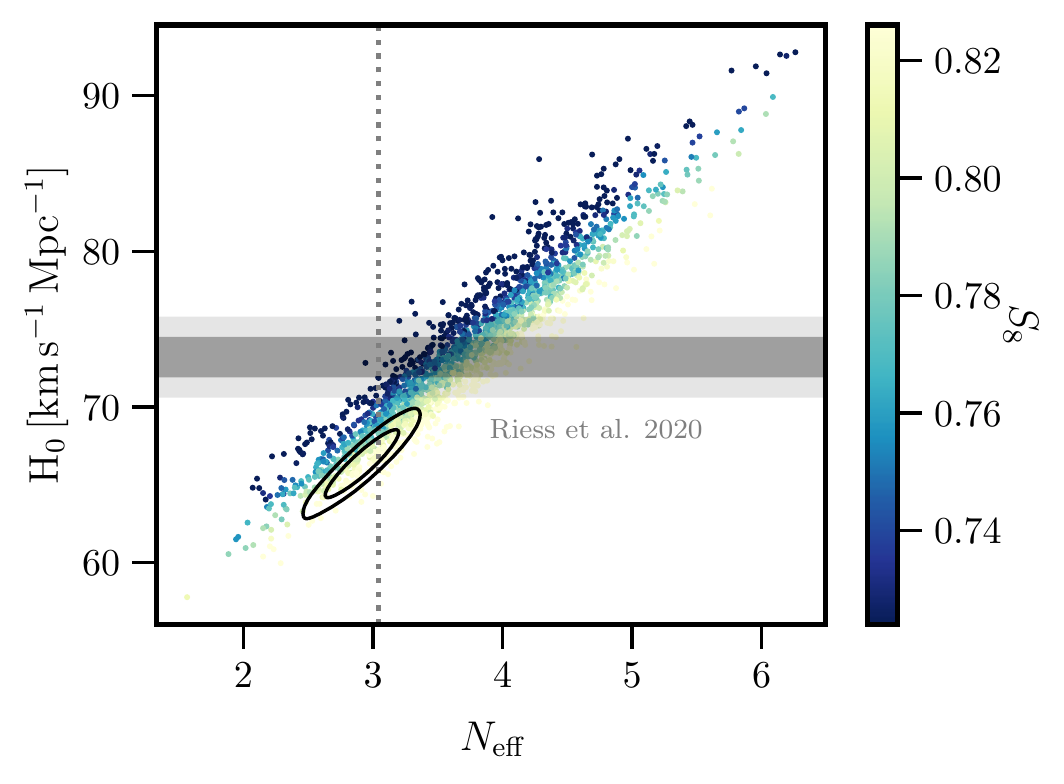}
  \end{subfigure}\hfill
  \begin{subfigure}{}
    \includegraphics[width=2.598in]{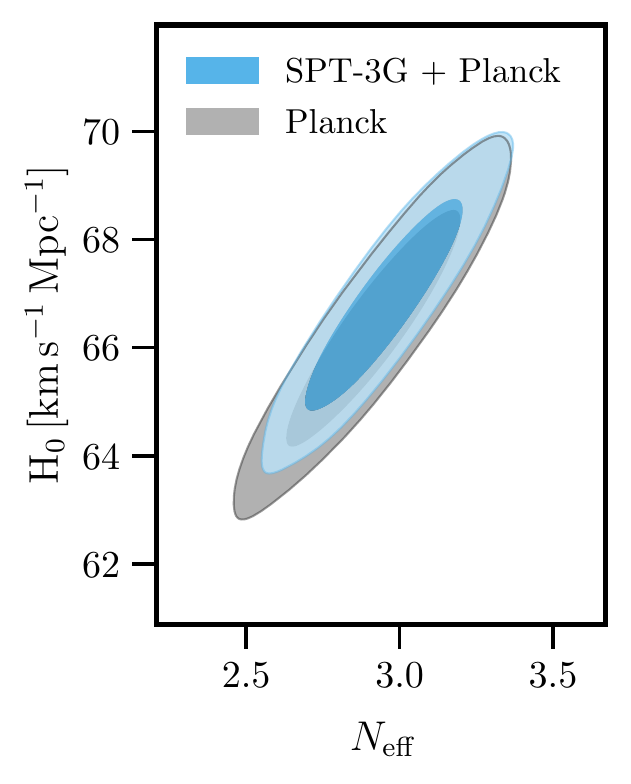}
  \end{subfigure}
  \caption{  \label{fig:neff}
\emph{Left panel:} We show samples in the $\Hubble$ vs. $\Neff$ plane from SPT-3G 2018 chains, colored according to $S_8$, a parameter describing the amplitude of matter perturbations today.
The color range has been chosen to match the $3\,\sigma$ range of the latest KiDS-1000 results \citep{heymans20}. 
For comparison, we also show the \planck{} 2D marginalized posterior probability (black lines), and the $2\,\sigma$ interval of the $\Hubble$ measurement from the distance-ladder of \citetalias{riess20}.
The dotted grey line is the standard model prediction of $\Neff = 3.044$.
\emph{Right panel:} Constraints from \planck{} (grey) by itself and jointly with SPT-3G 2018 (blue) in the $\Hubble$ vs. $\Neff$ plane for a \lcdm{}+$\Neff$ model. 
The inclusion of SPT-3G 2018 data tightens the constraint on $\Neff$ by $11\%$.
Given the high correlation between $\Hubble$ and $\Neff$, there is a similar refinement of the Hubble constant constraint.
Contours indicate the 68\% and 95\% probability regions.
}
\end{figure*}


The relativistic energy density in the early universe can be parametrized by $\Neff$, which is normalized to equal three for a thermal distribution of the three neutrino species in the standard model of particle physics. 
The expected value is $\Neff=3.044$, as there is a small non-thermal contribution to the neutrinos from electron-positron annihilation \citep{froustey20, bennett20}.\footnote{In our MCMC analysis we have assumed the standard model value of $\Neff=3.046$ based on \citet{snowmass13neutrinos}. However, this small change has a negligible impact on the results of this paper.}
There are a plethora of hypothesized particles that might change the observed $\Neff$, such as axion-like particles, hidden photons, gravitinos, or massless Goldstone bosons; the exact change in $\Neff$ depends on the nature of the particle and its coupling to the standard model \citep{abazajian16, pdg2020}.

We present constraints from SPT-3G 2018 data on \lcdm{}+$\Neff$ in Table \ref{tab:neff_param_table}. We find
\begin{equation}
\Neff = 3.70 \pm 0.70,
\end{equation}
which is within $0.9\,\sigma$ of the standard model prediction of 3.044.
As you can see in Figure \ref{fig:neff}, in CMB data constraints, higher values of $\Neff$ tend to lead to higher values of $\Hubble$; the slightly raised $\Neff$ value translates into a higher expansion rate, $\Hubble = 73.5 \pm 5.2\,\mathrm{km\,s^{-1}\,Mpc^{-1}}$.
While this is consistent with the distance-ladder measurement of $\Hubble$ by \citetalias{riess20} ($0.05\,\sigma$), the large uncertainty on the result means it is also consistent with CMB-based $\Hubble$ values in \lcdm{}.
As noted in Table \ref{tab:chisq}, this model barely changes the quality of fit compared to \lcdm{} ($\Delta\chi^2=-0.2$).

\begin{table}[tb]
\def\arraystretch{1.5}
\small
\setlength{\tabcolsep}{10pt}
\centering
\begin{tabular}{l c c}
\hline
 Model & $\Delta\chi^2$ & Additional d.o.f.\\
 \hline
 \lcdm{}+\Neff{} & -0.2 & 1\\
 \lcdm{}+\Yp{} & 0.1 & 1\\
 \lcdm{}+\Neff{}+\Yp{} & -1.8 & 2\\
 \lcdm{}+\mnu{} & 0.0 & 1\\
 \lcdm{}+\meff{}+\Neff{} & 0.1 & 2\\
 \lcdm{}+\curv{} & -0.3 & 1\\
\hline
\end{tabular}
\caption{
Improvement to the quality of fit for the cosmological models considered with respect to \lcdm{}, $\Delta\chi^2 = \chi^2_{\mathrm{\Lambda CDM}+} - \chi^2_{\mathrm{\Lambda CDM}}$.
We have run 10 minimisers without the annealer for each model and find that the $\chi^2$ of the best three runs typically span a range of the order of 0.1.
We also list the extra degrees of freedom (d.o.f.) added by each model extension compared to \lcdm{}.
}
\label{tab:chisq}
\end{table}
\begin{table}[tb]
\def\arraystretch{1.5}
\small
\setlength{\tabcolsep}{10pt}
\centering
\begin{tabular}{l c}
\hline
 Model & Volume Reduction\\
 \hline
 \lcdm{} & 1.5\\
 \lcdm{}+\Neff{} & 1.5\\
 \lcdm{}+\Yp{} & 1.4\\
 \lcdm{}+\Neff{}+\Yp{} & 1.7\\
 \lcdm{}+\mnu{} & 1.3\\
 \lcdm{}+\meff{} & 1.5\\
 \lcdm{}+\curv{} & 2.0\\
\hline
\end{tabular}
\caption{
The addition of the SPT-3G bandpowers to the \planck{} power spectra significantly reduces the 68\% confidence volume in parameter space for all extensions considered. 
As an approximate measure of the volume reduction, we report here the ratio of the square roots of the determinants of the parameter covariance matrices for \planck-only and \planck+SPT-3G. 
}
\label{tab:spt_improv}
\end{table}

The reported central value for $\Neff$ is consistent with, although higher than, the corresponding \planck{} and ACT DR4 values by $1.1\,\sigma$ and $1.7\,\sigma$, respectively. 
For the latter shift, we point out that our MCMC analysis of ACT DR4 yields $\Neff = 2.34 \pm 0.43$, which is less than the standard model prediction. 
The shift to lower \Neff{} compared to \lcdm{} in ACT DR4 is accompanied by shifts along the degeneracy directions in $\Omega_c h^2$ by -0.0097 and $n_s$ by -0.048. 
The constraints based on SPT-3G 2018 move in the opposite way along these same degeneracy axes, which places the central values of $\Omega_c h^2$ and $n_s$ 0.082 and 0.039 higher than in \lcdm{}, respectively, and $\Neff$ slightly above the standard model prediction.

Two tensions have been noted between \planck{} data and low-redshift measurements: in \lcdm{} one infers lower values of $\Hubble$ and higher values of $S_8\equiv\sigma_8\sqrt{\Omega_{\rm m}/0.3}$, a parameter describing the amplitude of matter perturbations today, from \planck{} data than from low-redshifts measurements \citep{planck18-6}.
The interplay between the inferred constraints from the SPT-3G 2018 bandpowers on $\Neff$, $\Hubble$, and $S_8$ is illustrated in the left panel of Figure \ref{fig:neff}.
$\Neff$ and $\Hubble$ are highly degenerate, such that an increase in $\Neff$ leads to higher values of $\Hubble$. 
The SPT-3G data alone allow high values of \Neff{} and correspondingly high values of \Hubble{} that overlap with the distance-ladder measurement in \citetalias{riess20} (the horizontal grey bands).
However, such high values of \Neff{} and \Hubble{} are ruled out by the \planck{} data (black contours), so the tension persists, although at lower significance due to the larger uncertainty on \Hubble{} when varying \Neff{} for CMB data.
The $S_8$ value for each sample in the SPT-3G chains is represented by the color, with the color range chosen to represent the 3\,$\sigma$ range of the cosmic shear analysis by \citet{heymans20}. 
Notably, $S_8$ varies perpendicular to the main degeneracy direction in the data, thus allowing \Neff{} to vary does little to reduce the tension in constraints of $S_8$.

The right panel of Figure \ref{fig:neff} shows the constraints on $\Neff$ and $\Hubble$ from the SPT-3G 2018 and \planck{} data.
The full results are listed in Table \ref{tab:neff_param_table}.
In particular, the joint constraint on the effective number of neutrino species is
\begin{equation}
\Neff = 2.95 \pm 0.17,
\end{equation}
which is within $0.6\,\sigma$ of the standard model prediction. 
Adding the SPT-3G 2018 bandpowers to the \planck{} data tightens the $\Neff$ and $\Hubble$ constraints by $11\%$ and reduces the square-root of the determinants of the parameter covariance matrices in this 7-parameter model by a factor of $1.5$ (see Table \ref{tab:spt_improv}).


\subsection{Primordial Helium Abundance, $\boldsymbol{Y_{P}}$}
\label{subsec:Yp}

The primordial helium abundance is a direct measure of the equilibrium abundance of neutrons relative to protons during BBN, when the reactions that interconvert them become slow compared to the expansion rate. Virtually all neutrons end up in helium atoms during this period.
The equilibrium abundance when these reactions freeze out depends on all known forces and as such measurements of the primordial helium abundance are a powerful probe of our understanding of particle physics.

The CMB anisotropies are sensitive to the helium abundance because helium's first electron has a higher binding energy than hydrogen's, which means that the helium recombination happens earlier than hydrogen.
As a consequence, increasing the helium abundance lowers the free electron density during hydrogen recombination.
The presence of fewer free electrons reduces the likelihood for Thomson scattering.
The photon mean-free path is increased, leading the CMB power spectra at high $\ell$ to be suppressed as structure on small scales is washed out.
Therefore, CMB power spectrum measurements can leverage the change in the Silk damping scale to constrain $\Yp$.

The constraints from the SPT-3G 2018 bandpowers on \lcdm{}+$\Yp$ are given in Table \ref{tab:neff_param_table}.
We find
\begin{equation}
\Yp = 0.225 \pm 0.052,
\end{equation}
which is consistent with the BBN prediction of 0.2454 at $0.4\,\sigma$ \citepalias{dutcher21}.
The SPT-3G 2018 helium constraint is also consistent with the latest CMB results from \planck{} ($0.3\,\sigma$, \citep{planck18-6}) and ACT DR4 ($0.5\,\sigma$, \citep{aiola20}), as well as recent measurements of H\,II regions of metal-poor galaxies ($0.4\,\sigma$, \citep{aver20}).
Current measurements of the primordial helium abundance are consistent with BBN expectations.
The change to the quality of fit for this model compared to \lcdm{} is insignificant ($\Delta\chi^2=0.1$, see Table \ref{tab:chisq}).


We look at joint constraints from SPT-3G 2018 and \planck{} (see Table \ref{tab:neff_param_table}). 
As noted in Table \ref{tab:spt_improv}, the addition of SPT-3G 2018 data to the \planck{} data reduces the square-root of the determinants of the parameter covariance matrices in this 7-parameter model by a factor of $1.4$.
The measurement of $\Hubble$ is improved by $8\%$, while the uncertainty on the helium fraction is essentially unchanged, yielding
\begin{equation}
\Yp = 0.234 \pm 0.012.
\end{equation}
This measurement is consistent with the BBN prediction of 0.2454 (note the BBN prediction varies with the \lcdm{} parameters) at $0.9\,\sigma$, as well as the H II region-based measurement of \citet{aver20} (0.9\,$\sigma$).


\subsection{Effective Number of Neutrino Species and Primordial Helium Abundance, $\boldsymbol{N_{\rm eff} + Y_{P}}$}
\label{subsec:Neff+Yp}

We now look at the constraints when simultaneously varying $\Neff$ and $\Yp$. 
Since BBN makes precise predictions for the primordial helium abundance as a function of the effective number of neutrino species and other parameters, the constraint on $\Neff$ in \S\ref{subsec:Neff} implicitly assumes that any extra relativistic species are present during both BBN and recombination. 
Simultaneously varying $\Neff$ and $\Yp$ removes this assumption and allows for independent constraints on the relativistic energy density during each epoch. 

\begin{figure*}[ht]
  \centering
  \begin{subfigure}{}
    \includegraphics[width=4.33in]{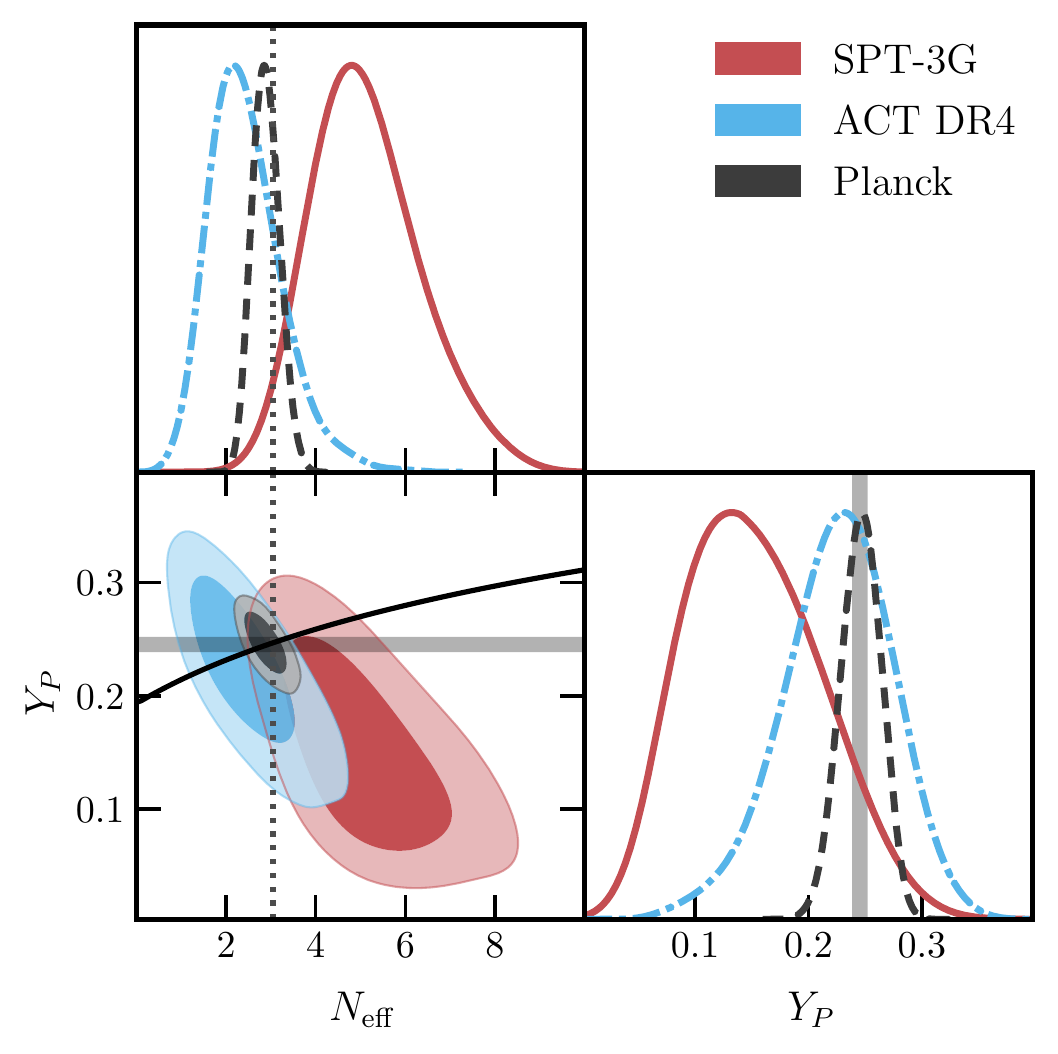}
  \end{subfigure}\hfill
  \begin{subfigure}{}
    \includegraphics[width=2.598in]{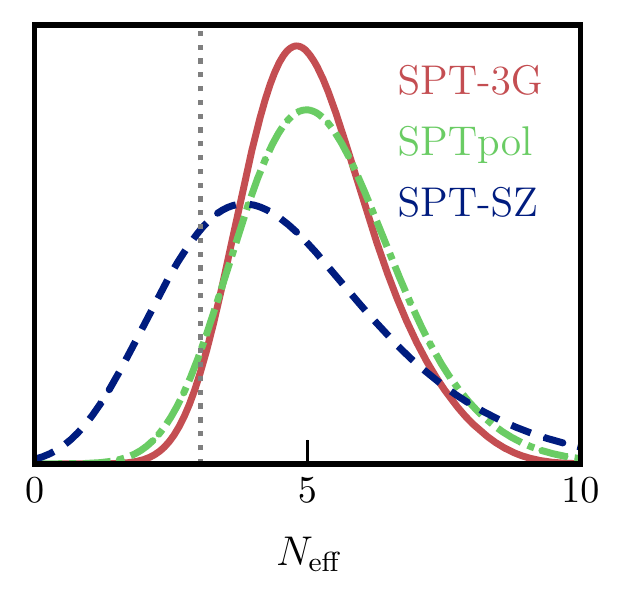}
  \end{subfigure}
  \caption{
\emph{Left:}
Constraints on $\Neff$ and $\Yp$. The contours indicating the 68\% and 95\% probability regions inferred from the SPT-3G 2018, \planck{}, and ACT DR4 datasets are shown in red (solid), dark grey (dashed), and blue (dash-dotted), respectively. The vertical dotted grey line indicates the standard model prediction $\Neff=3.044$. The solid black line in the lower left panel shows the BBN prediction for the primordial helium abundance while the light grey band in panels with \Yp{} shows the 95\% confidence interval of the latest H\,II region-based measurement \citep{aver20}.  
\emph{Right:} Successive generations of SPT observations have improved constraints on \Neff, with SPT-3G 2018 achieving a $57\%$ and $15\%$ improvement over SPT-SZ and SPTpol, respectively. 
The lines show the marginalized 1D posteriors for $\Neff$ in the \lcdm{}+$\Neff$+$\Yp$ model from SPT-3G 2018 (red, solid), SPTpol (green, dash-dotted), and SPT-SZ data (blue, dashed).
  }
  \label{fig:neff_yp}
\end{figure*}

We present the constraints SPT-3G 2018 places on \lcdm{}+$\Neff$+$\Yp$ in Table \ref{tab:neff_param_table} and show the marginalized 1D and 2D posterior probabilities for \Neff{} and \Yp{} in the left panel of Figure \ref{fig:neff_yp}.
We find
\begin{equation}
\begin{split}
\Neff &= 5.1 \pm 1.2,\\
\Yp &= 0.151 \pm 0.060.
\end{split}
\end{equation}
The central value of $\Neff$ is $1.7\,\sigma$ higher than the standard model prediction of 3.044, while the $\Yp$ value is $1.6\,\sigma$ lower than the \lcdm{} prediction of 0.2454; the parameters shift along the degeneracy direction in the $\Neff$ vs. $\Yp$ plane as shown in the left panel of Figure \ref{fig:neff_yp}.
The plot also shows that consistency with BBN, as well as departures to $\Yp$ values far below the BBN expectation, are compatible with the SPT-3G data.
The fit quality improves by only $\Delta\chi^2=-1.8$ compared to \lcdm{} for two additional parameters (see Table \ref{tab:chisq}).
The mild preference is driven by the data at $\ell < 800$; removing the lower multipoles shifts the best-fit values towards the \lcdm{} expectations.
These angular-scales have been well-measured by \planck{}, which does not share this trend.
Similar to \S \ref{subsec:Neff}, we find that the shifts in the values of $\Neff$ and $\Yp$ lead to increases in $\Omega_c h^2$ and $n_s$ by 0.026 and 0.020 compared to \lcdm{}, respectively.

The left panel of Figure \ref{fig:neff_yp} compares the posteriors in the $\Neff$ vs. $\Yp$ plane from SPT-3G 2018, \planck{}, and ACT DR4. 
As should be expected, all three show a similar degeneracy axis, where increasing $\Neff$ decreases $\Yp$.
The central value of the SPT-3G 2018 constraint is higher along the $\Neff$ axis (and lower along the $\Yp$ axis) than \planck{}, which in turn is higher than ACT DR4.
Our central value of $\Neff$ is $1.8\,\sigma$ higher than the \planck{} value, and larger than the ACT DR4 value by the same amount (although it is lower than \planck{}, its associated uncertainty is larger).
The $\Yp$ value from SPT-3G is lower than the \planck{} and ACT DR4 ones by $1.5\,\sigma$ and $1.0\,\sigma$, respectively.

To quantify the agreement between SPT-3G 2018 and \planck{} in the full parameter space, we calculate the $\chi^2$ of the differences in the mean values of the parameters using the inverse of the sum of parameter covariance matrices.
We use a combined parameter, $10^9 A_s e^{-2\tau}$, to account for the \planck{}-based $\tau$ prior used in the SPT-3G constraints.
Thus the comparison covers seven parameters $(\Omega_b h^2, \Omega_c h^2, \theta_{MC}, 10^9 A_s e^{-2\tau}, n_s, \Neff, \Yp)$.
We find $\chi^2=12.3$ between the SPT-3G 2018 and \planck{} datasets, which corresponds to a probability to exceed (PTE) of 9\%. This is within the central 95\% confidence interval $[2.5\%,97.5\%]$ and we conclude that the two datasets are consistent with one another.

The same comparison for SPT-3G 2018 and ACT DR4 yields $\chi^2=17.8$, which translates to a PTE of 1\%.
This low PTE is driven by differences in the preferred baryon density. 
The $\Omega_b h^2$ value for ACT DR4 is $2.6\,\sigma$ below the SPT-3G 2018 result. 
The low baryon density inferred from ACT DR4 has been previously noted by \citet{aiola20}, who explain that the shift is related to degeneracies over the limited angular multipole range probed.
Removing $\Omega_b h^2$ from the comparison reduces the $\chi^2$ to $12.7$ and raises the PTE to $5\%$. 
Outside of the noted variation in the preferred baryon density with ACT DR4, we conclude that the parameter constraints in the \lcdm{}+$\Neff$+$\Yp$ model are consistent across  the three experiments.
 
The SPT-3G 2018 primordial helium abundance constraint is $1.6\,\sigma$ lower than the most precise measurement based on the H II regions of metal-poor galaxies \citep{aver20}.
While the SPT-3G 2018 data alone allow for very high expansion rates in the \lcdm{}+$\Neff$+$\Yp$ model extension, $H_0 = 80.4 \pm 7.2\,\mathrm{km\,s^{-1}\,Mpc^{-1}}$, the addition of \planck{} data significantly tightens the $\Hubble$ constraint and pulls the value down to $H_0 = 67.7 \pm 1.8\,\mathrm{km\,s^{-1}\,Mpc^{-1}}$. 
We discuss the results with \planck{} in more detail below.

\skipt{
We compare the constraints SPT-3G 2018, ACT DR4, and \planck{} place on \lcdm{} extension parameters $\Neff$ and $Y_P$ in Figure \ref{fig:neff_yp_fluct}.

\begin{figure}[h!]
\includegraphics[width=8.6cm]{figures/neff_yp_fluct_zoom.pdf}
\centering
\caption{
Constraints using SPT-3G 2018, \planck{} and \planck{} data on $\Neff$ (top panel) and the primordial helium abundance, $\Yp$ (bottom panel). The blue (left, round) markers show results from varying either parameter individually, while the orange (right, square) markers correspond to a model which allows both parameters to vary simultaneously. The dashed grey line in the top panel indicates the standard model prediction of $\Neff = 3.044$, while the grey band in the bottom panel shows the $2\,\sigma$ confidence interval of the latest H II region-based measurement of the primordial helium abundance \citep{aver20}.
}
\label{fig:neff_yp_fluct}
\end{figure}
}

Comparison in the \lcdm{}+$\Neff$+$\Yp$ model shows the improvement across successive SPT power spectrum measurements. 
We compile the 1D marginalized posterior for $\Neff$ as constrained by SPT-SZ, SPTpol, and SPT-3G 2018 for this two-parameter extension in the right panel of Figure \ref{fig:neff_yp}. 
Across three generations of experiments from SPT-SZ to SPTpol to SPT-3G 2018, the uncertainty on the effective number of neutrino species has shrunk from $\sigma(\Neff) = 1.9$ to 1.4 to 1.2.
Furthermore, we note that the SPT-SZ and SPTpol datasets were based on nearly complete multi-year surveys, whereas the SPT-3G 2018 data was recorded over a four-month period (half of a typical observing season) and data is still being collected. 

\begin{figure*}[htp!]
  \centering
  \begin{subfigure}{}
    \includegraphics[width=3.464in]{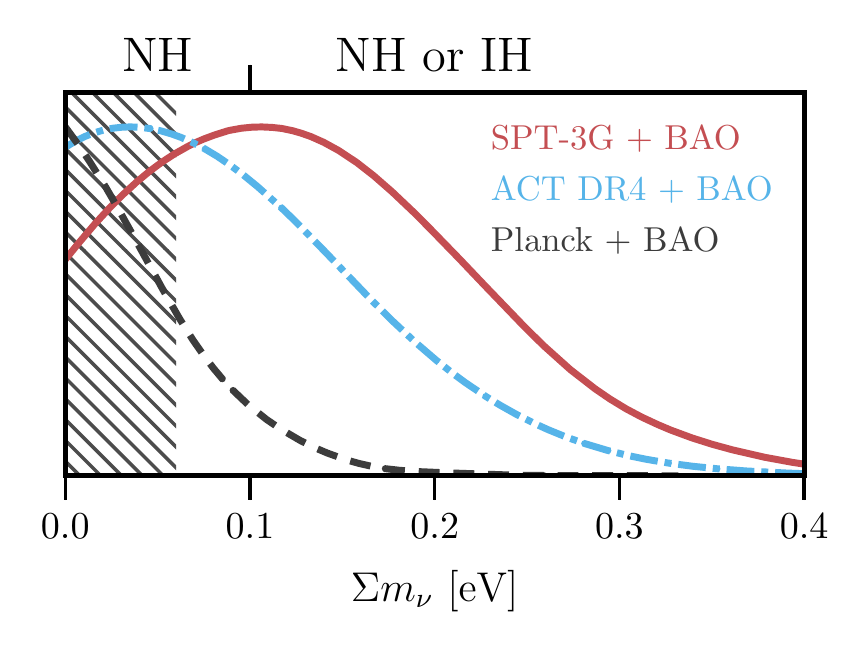}
  \end{subfigure}\hfill
  \begin{subfigure}{}
    \includegraphics[width=3.464in]{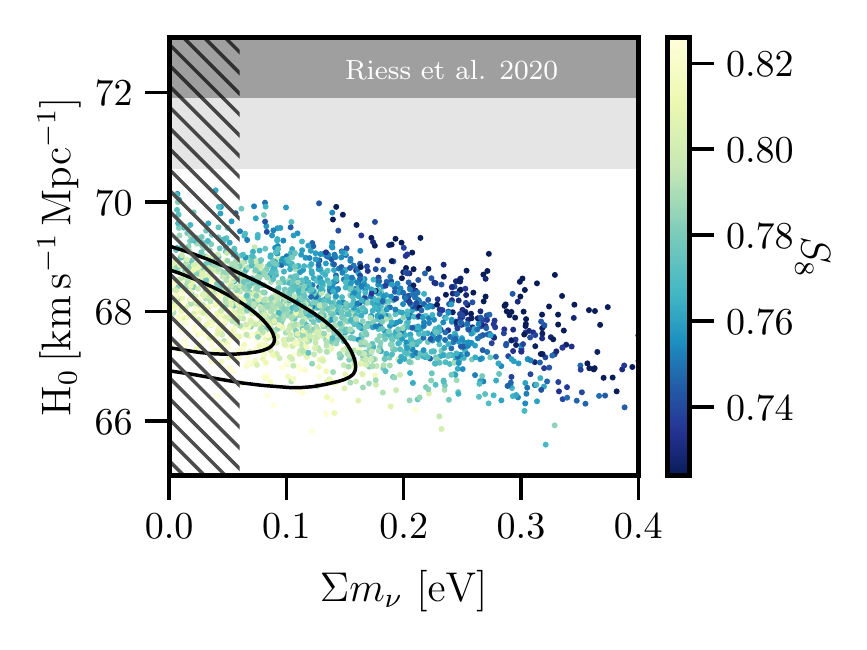}
  \end{subfigure}
  \caption{
\emph{Left:} The CMB and BAO data place upper limits on the sum of neutrino masses, $\mnu$. The results from combining BAO data with SPT-3G 2018, ACT DR4, and \planck{} are shown in red (solid), black (dashed), and blue (dash-dotted), respectively. The hatched region is ruled out by neutrino oscillation observations, which require $\mnu > 0.06\,\mathrm{eV}$ in the normal hierarchy and $\mnu > 0.1\,\mathrm{eV}$ in the inverted hierarchy. 
The allowed mass-ranges of the normal hierarchy (NH) and inverted hierarchy (IH) are also marked on the top of the plot. 
\emph{Right:} Lower neutrino masses are correlated with higher values of the Hubble constant. 
The colored points show values for $\Hubble$ from samples from the SPT-3G 2018 + BAO chains. 
The color represents $S_8$ for that chain sample, with the color scale chosen to cover the $3\,\sigma$ band of the latest KiDS-1000 results \citep{heymans20}. 
The black lines show the 2D marginalized 68\% and 95\% posterior probability from \planck{}. 
The dark (light) grey region corresponds to the $1\,\sigma$ ($2\,\sigma$) band for the \citetalias{riess20} distance-ladder Hubble measurement. 
As in the left panel, the hatched region indicates the mass range ruled out by neutrino oscillation observations.
}
\label{fig:mnu}
\end{figure*}

Joint constraints from SPT-3G 2018 and \planck{} are given in Table \ref{tab:neff_param_table}. 
Adding the SPT-3G to \planck{} data reduces the square-root of the determinants of the parameter covariance matrices in this 8-parameter model by a factor of $1.7$ (see Table \ref{tab:spt_improv}), signalling a substantial reduction in the allowed parameter volume. 
For SPT-3G 2018 and \planck{}, we report
\begin{equation}
\begin{split}
\Neff &= 3.13 \pm 0.30,\\
\Yp &= 0.230 \pm 0.017.
\end{split}
\end{equation}
These values are offset from their standard model predictions by $0.3\,\sigma$ and $0.1\,\sigma$, respectively. 
The mean of the helium fraction posterior is $0.7\,\sigma$ less than the H\,II region-based measurement of \citet{aver20}.

\subsection{Neutrino Masses, $\boldsymbol{\mnu}$}
\label{subsec:mnu}

\begin{figure}[ht]
\includegraphics[width=8.6cm]{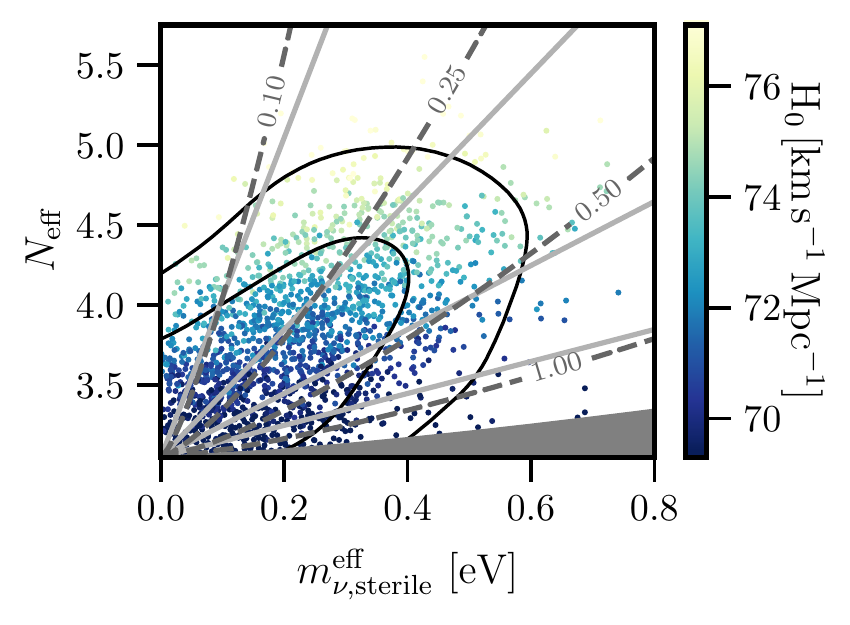}
\centering
\caption{
The SPT-3G 2018 and BAO constrain the energy density and effective mass of a sterile neutrino; higher values of $\Neff$ tend to correlate with higher values of $\meff$.
The colored points show the values of \Neff{} and \meff{} from samples in the SPT-3G 2018 + BAO chains, with the color determined by each sample's \Hubble{} value. 
The color scale is chosen to cover the $3\,\sigma$ range of the \citetalias{riess20} distance-ladder result.
The black lines denote the 2D marginalized 68\% and 95\% probability regions for these data.
The dark grey dashed lines and light grey solid lines correspond to a constant physical mass of $0.1, 0.25, 0.5, 1\,\mathrm{eV}$ (clockwise) assuming a thermal distribution of the sterile neutrino momenta and the Dodelson-Widrow mechanism \cite{dodelson94}, respectively.
The solid grey region is excluded by the prior $m^{\rm thermal}_{\rm sterile}<2\,\mathrm{eV}$.
}
\label{fig:neff_meff_H0}
\end{figure}

The neutrino sector is one of the least understood areas of the standard model of particle physics.
Determining neutrino hierarchy and the mechanism by which neutrinos attain their mass are key questions. 
CMB observations allow us to constrain the sum of neutrino masses, $\mnu$, and are complementary to terrestrial experiments, which have so far measured the squared mass splittings and the sign of one splitting \citep{lesgourgues14, dolinski19, acero19}.


We present the constraints on \lcdm{}+$\mnu$ placed by SPT-3G 2018 alone and in combination with BAO and \planck{} data in Table \ref{tab:mnu_param_table}. SPT-3G 2018 alone constrains $\mnu$ to $0.69 \pm 0.67\,\mathrm{eV}$, with an upper limit of $\mnu<2.0\,\mathrm{eV}$ at 95\% confidence.
We report no change to the quality of fit for this model compared to \lcdm{} (see Table \ref{tab:chisq}).

We add BAO measurements to improve the $\mnu$ constraint. 
The low-redshift BAO points significantly reduce the large degeneracy between the expansion rate today and sum of the neutrino masses that exists in the SPT-3G data alone; the uncertainty on \Hubble{} drops from 5.3 to 0.70\,km s$^{-1}$ Mpc$^{-1}$ as can be seen in columns 1 and 3 of Table \ref{tab:mnu_param_table}. 
The upper limit from on $\mnu$ SPT-3G plus BAO is:
\begin{equation}
\mnu < 0.30\,\mathrm{eV}\,(\mathrm{95\%\,CL}).
\end{equation}
This limit is weaker than the 95\% CL upper limits of $0.13\,\mathrm{eV}$ and $0.24\,\mathrm{eV}$ set by \planck{} and ACT DR4 in combination with BAO measurements, respectively. 
We show the associated marginalized 1D posteriors for all three datasets in the left panel of Figure \ref{fig:mnu}.
As can be seen there, some of the difference in the upper limits is due to where the posteriors peak, with the SPT-3G posterior reaching its maximum at $\sim 0.11\,\mathrm{eV}$.

We highlight the interplay between the joint constraints from SPT-3G 2018 and BAO data on the sum of the neutrino masses $\mnu$, Hubble constant $\Hubble$, and a parameter describing the amplitude of density perturbations today, $S_8$, in the right panel of Figure \ref{fig:mnu}.
Massive neutrinos offer no resolution to the Hubble tension: increasing the neutrino mass lowers the expansion rate inferred from the CMB and increases the gap between early- and late-time probes.
The combination of the SPT-3G 2018 and BAO datasets rules out $\Hubble > 70$\,km s$^{-1}$ Mpc$^{-1}$ at $2.9\,\sigma$, leaving a $3.5\,\sigma$ rift to the most recent distance-ladder measurement by \citetalias{riess20} (indicated in grey in the figure).
It is interesting to note that in the $\mnu$, $\Hubble$, $S_8$ space shown, the measurements of \citetalias{riess20} and \citet{heymans20} lie in the same direction relative to the \planck{} constraints; increasing the value of $\Hubble$ at fixed $\mnu$ also decreases the inferred $S_8$ value, thus improving the consistency with the local measurements of $\Hubble$ from \citetalias{riess20} and of $S_8$ from \citet{heymans20}.

The parameter constraints from combining SPT-3G 2018, \planck{}, and BAO data on \lcdm{}+$\mnu$ are shown in Table \ref{tab:mnu_param_table}.
The addition of \planck{} power spectrum data reduces the upper limit on \mnu{} by more than a factor of two to:
\begin{equation}
\mnu < 0.13\,\mathrm{eV}\,(\mathrm{95\%\,CL}).
\end{equation}
The \planck{} large-scale temperature data adds information from both the late time integrated Sachs-Wolfe effect and the observed peak smoothing, which depends on the amount of gravitational lensing. 
Previous works have noted that one reason the \planck{} data favor low neutrino masses is the excess peak-smoothing observed in the \planck{} $\TT$ bandpowers \citep{planck18-6, bianchini20}.
Removing the \planck{} TT bandpowers (keeping \planck{} TE and EE) from the data combination relaxes the upper limit by 50\% to $\mnu<0.20\,\mathrm{eV}$.
As an approximate estimate of how much information is added by the SPT-3G data, we calculate the ratio for the square-root of the determinants of the parameter covariance matrices when adding the SPT-3G 2018 dataset to \planck{} (including the $\TT$ bandpowers) and BAO data to be 1.3 (see Table \ref{tab:spt_improv}). 
Adding the SPT-3G data to the \planck{} and BAO data thus substantially reduces the allowed parameter volume. 

\subsection{Sterile Neutrinos, $\boldsymbol{\meff}$}
\label{subsec:meff}

Sterile neutrinos are a hypothesized species of neutrinos that do not interact through the weak force, only gravitationally.
We investigate the model formulated by the \planck{} collaboration, which we describe briefly here (for more details see \citet{planck13-16, planck15-13, planck18-6}).
Motivated by the results of \citet{acero19}, we assume minimal neutrino masses in the normal mass hierarchy, which we approximate as two massless and one massive active neutrino with a mass of $0.06\,\mathrm{eV}$. To these we add one massive sterile neutrino with an abundance and distribution across momentum arising from its mixing with active neutrinos.

We consider both a thermal distribution and, as in the Dodelson-Widrow (DW) mechanism \citep{dodelson94}, a distribution proportional to that of the active neutrinos with a scaling factor dependent on the mixing angle between the active and sterile neutrinos.
Since the two scenarios are cosmologically equivalent, we sample over the effective mass $\meff = 94.1\,\Omega_{\mathrm{\nu, sterile}}h^2$~eV, which maps to the physical mass according to $m^{\rm physical}_{\rm sterile} = \meff (\Delta\Neff)^{\alpha}$, where $\Delta\Neff$ is the deviation of the effective number of neutrino species from the standard model prediction, and $\alpha = -3/4$ for a thermal distribution of sterile neutrino momenta or $\alpha = -1$ for the DW mechanism.

Sterile neutrinos with physical masses $\gtrsim 10\,\mathrm{eV}$ become non-relativistic well before recombination and, depending on their mass, mimic warm or cold dark matter. 
To avoid this regime, we focus our analysis on the region in $(\Neff, \meff)$ space that corresponds to a physical mass of $m^{\rm physical}_{\rm sterile} <2\,\mathrm{eV}$, assuming a thermal distribution of sterile neutrino momenta.\footnote{The results only change slightly if we assume the DW scenario for this prior instead of a thermal distribution of sterile neutrino momenta.}
Since sterile neutrinos in this region of parameter space would be relativistic at last-scattering, we would expect them to increase $\Neff$.


We present the constraints the SPT-3G 2018 dataset places by itself and in combination with BAO on \lcdm{}+$\meff$ in Table \ref{tab:meff_param_table}. 
The SPT-3G 2018 dataset is consistent with the null hypothesis of no sterile neutrinos, constraining $\Delta\Neff < 1.8$ and $\meff < 1.5\,\mathrm{eV}$ at 95$\%$ confidence. 
Including BAO data tightens these 95\% CL limits to:
\begin{equation}
\begin{split}
\Delta\Neff &< 1.6,\\
\meff &< 0.50\,\mathrm{eV}.
\end{split}
\end{equation}
As noted in Table \ref{tab:chisq}, we find that the quality of fit for this model does not change significantly from \lcdm{} ($\Delta\chi^2=0.1$).
The \planck{} and ACT DR4 datasets also yield no evidence for sterile neutrinos: in combination with BAO data we infer $\Delta\Neff < 0.29\,\meff<0.24,\,\mathrm{eV}$ from \planck{} and $\Delta\Neff < 0.58,\,\meff<0.32\,\mathrm{eV}$ from ACT DR4.

We plot the constraints placed by SPT-3G 2018 + BAO in the $\Neff$ vs. $\meff$ plane in Figure \ref{fig:neff_meff_H0}, where the degeneracy of these parameters with $\Hubble$ can be observed.
We report $\Hubble = 71.6 \pm 2.2$\,km s$^{-1}$ Mpc$^{-1}$, which is higher than the \lcdm{} value due to the increase in the effective number of neutrino species, similar to \S\ref{subsec:Neff}.
While an increase to $\Neff$ of the size needed to reconcile late- and early-time probes of $\Hubble$ is allowed by the SPT-3G 2018 dataset, it is disfavored by \planck{} \citep{planck18-6}.

Joint constraints from SPT-3G 2018, \planck{}, and BAO data on sterile neutrinos are given in Table \ref{tab:meff_param_table}. 
We find 95\% CL upper limits of
\begin{equation}
\begin{split}
\Delta\Neff &< 0.30,\\
\meff &< 0.20\,\mathrm{eV}.
\end{split}
\end{equation}
The addition of \planck{} data reduces the upper limit on $\Neff$ five-fold, and as a result tightens the posterior on \Hubble{} to $68.30 \pm 0.70$\,km s$^{-1}$ Mpc$^{-1}$. 
The CMB-preferred value of \Hubble{} remains in tension with the distance-ladder measurement of \citetalias{riess20} at $3.5\,\sigma$. 
Finally, as an indicator of the extent to which SPT-3G data reduces the allowed parameter volume in the 8-dimensional space, we once again calculate the square-root of the determinants of the parameter covariance matrices, finding a reduction by a factor of 1.6 when adding the SPT-3G 2018 dataset to \planck{} and BAO data.

\subsection{Spatial Curvature, $\boldsymbol{\curv}$}
\label{subsec:curv}

Inflation in the early universe should suppress any primordial spatial curvature, leading to a flat universe today to well below the precision of current measurements. 
While primary CMB observations can test this assumption, they suffer from geometric degeneracies which limit their precision.
The \planck{} dataset prominently gives support for a closed universe at well over $2\,\sigma$ when considering primary CMB data alone.
However, adding CMB lensing or BAO data drives the posterior back to $\curv =0$ \citep{planck18-6}.

\begin{figure}[t!]
\includegraphics[width=8.6cm]{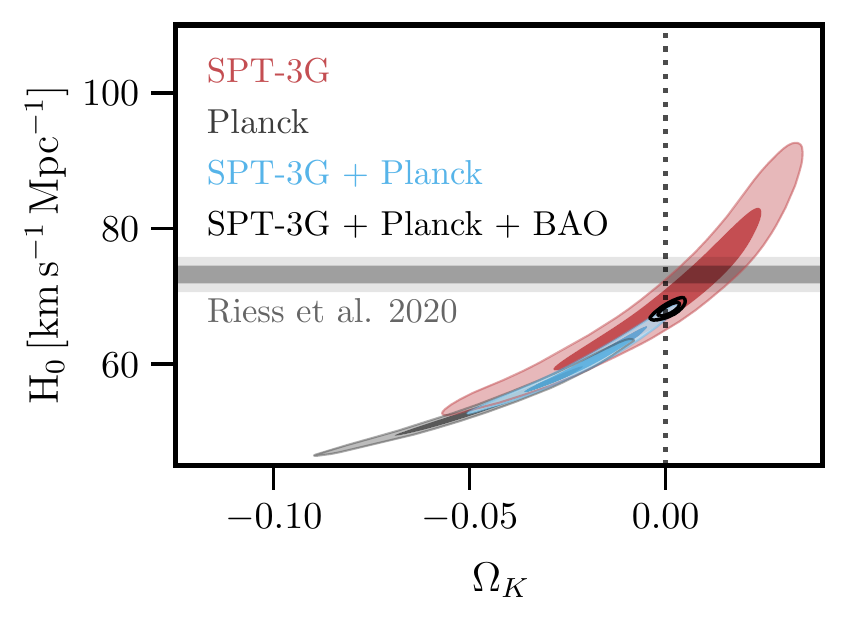}
\centering
\caption{
Marginalized 2D 68\% and 95\% posterior probability contours in the $\Hubble$ vs. $\curv$ plane for SPT-3G (red), \planck{} (dark grey), SPT-3G+\planck{} (blue), and the combination of SPT-3G 2018, \planck{}, and BAO data (black lines). 
The SPT-3G data by itself places constraints competitive with \planck{} on curvature, in part due to the upturn in the degeneracy between \curv{} and \Hubble{} as \curv{} increases. 
The combined SPT-3G 2018 and \planck{} data results in a curvature constraint consistent with the standard model prediction at $1.8\,\sigma$.
While this raises the inferred $\Hubble$ value compared to \planck{}-only constraints to $60.6 \pm 3.4$\,km s$^{-1}$ Mpc$^{-1}$, it remains in tension with the distance-ladder measurement by \citetalias{riess20}, for which we show the $2\,\sigma$ interval in the horizontal grey bands, at $3.5\,\sigma$
}
\label{fig:curv}
\end{figure}

We report constraints on \lcdm{}+$\curv$ from SPT-3G 2018 alone and jointly with BAO data in Table \ref{tab:curv_param_table}. 
From SPT-3G 2018 alone we determine $\curv = 0.001^{+0.018}_{-0.019}$. 
This is perfectly consistent with a flat universe. 
We highlight that the marginalized confidence interval for $\curv$ is close to the precision of the \planck{} data ($\curv = -0.044^{+0.018}_{-0.015}$).
The precision of this result is not simply a reflection of the quality of the SPT-3G 2018 dataset, but also due to increasing slope of the degeneracy between \Hubble{} and \curv{} observable in Figure \ref{fig:curv}.
This model barely changes the quality of fit compared to \lcdm{} ($\Delta\chi^2=-0.3$, see Table \ref{tab:chisq}).

With the primary CMB information alone, spatial curvature is degenerate with the Hubble constant; the geometric impact of an open universe on the distance to the last-scattering surface can be compensated for by a higher expansion rate. Adding BAO information breaks this degeneracy, and for SPT-3G 2018 plus BAO data we report 
\begin{equation}
\curv = -0.0014 \pm 0.0037.
\end{equation}
The central value is consistent with flatness at $0.4\,\sigma$.
The BAO data also reduces the error on the $\Hubble$ determination from $\sigma(\Hubble) = 8.5$\,km s$^{-1}$ Mpc$^{-1}$ by a factor of $11$ to $\sigma(\Hubble) = 0.76$\,km s$^{-1}$ Mpc$^{-1}$ for the SPT-3G 2018 dataset. The combination of SPT-3G 2018 and BAO data constrains $\Hubble$ to $68.11 \pm 0.76$\,km s$^{-1}$ Mpc$^{-1}$. 
 Given the inferred curvature is nearly zero, it is unsurprising that the $\Hubble$ central value is basically unchanged from the result in the standard 6-parameter flat \lcdm{} model. 
The mean value of $\Hubble$ is $3.4\,\sigma$ lower than the \citetalias{riess20} distance-ladder measurement.

The SPT-3G 2018 and \planck{} parameter posteriors are statistically consistent in the \lcdm{}+$\curv$ model. 
We compute the parameter-level $\chi^2$ between the two datasets across the six free cosmological parameters as in \S\ref{subsec:Neff+Yp} and find $\chi^2 = 13.0$ (PTE = $4.3\%$). 
The largest differences are in $\curv$ and $\theta_{MC}$, which are degenerate with one another and offset along this degeneracy direction by $1.8\,\sigma$ in both parameters.
However, we point out again that, as illustrated by the curved ellipses in Figure \ref{fig:curv}, the posteriors on these parameters are not well-described by a simple N-dimensional Gaussian assumed in a covariance matrix formalism.
Therefore, this result only provides a qualitative view of the more complex parameter space.

We combine the SPT-3G 2018 and \planck{} data, reporting joint parameter constraints in Table \ref{tab:curv_param_table}. 
The interplay of the different datasets is illustrated in Figure \ref{fig:curv}. 
We find that the inclusion of SPT data pulls the inferred curvature value towards flatness: $\curv = -0.020 \pm 0.011$. 
The \curv{} constraint is refined by $56\%$ compared to the \planck{} result and its central value is within $1.8\,\sigma$ of the standard model prediction of zero. 
This large improvement is in part owed to the aforementioned offset in the $\curv$ vs. $\theta_{MC}$ plane between the individual constraints from SPT-3G 2018 and \planck{} and the shift of the constraint in the highly non-Gaussian parameter space.
We approximate the reduction in the allowed parameter volume by again looking at the ratio of the square-root of the determinants of the parameter covariance matrices when adding the SPT-3G 2018 dataset to \planck{}, finding a ratio of 2.0.  
As can be seen in Table \ref{tab:spt_improv}, this extension shows the largest improvement from the SPT-3G data. 
The joint constraint on $\Hubble$ is $60.6 \pm 3.4\,\mathrm{km\,s^{-1}\,Mpc^{-1}}$, which is $3.5\,\sigma$ lower than the distance-ladder measurement by \citetalias{riess20}.

Combining the two CMB datasets with BAO information yields
\begin{equation}
\curv = 0.0009 \pm 0.0018,
\end{equation}
which is consistent with flatness ($0.5\,\sigma$). The addition of BAO data also tightens the $\Hubble$ constraint to $68.05 \pm 0.67\,\mathrm{km\,s^{-1}\,Mpc^{-1}}$. This value is in tension with the latest distance-ladder measurement at $3.5\,\sigma$.


\section{$\boldsymbol{\Hubble}$ from Temperature and Polarization Data}
\label{sec:t_vs_p}

\begin{figure*}[ht!]
\includegraphics[width=17.2cm]{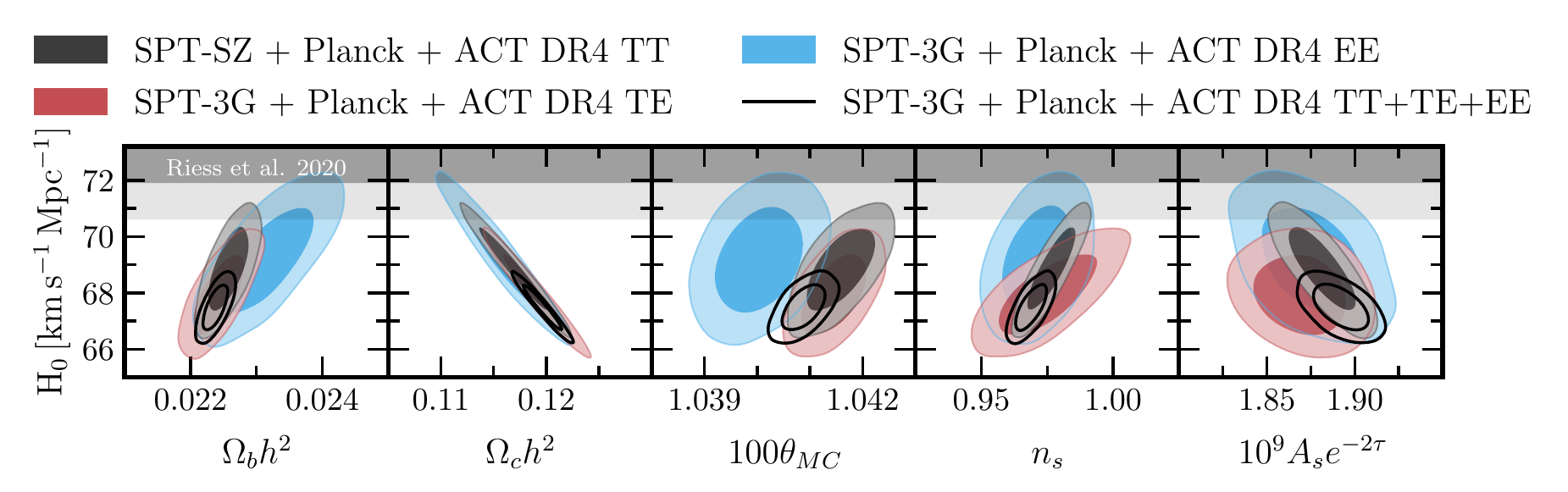}
\centering
\caption{
Comparison of the 2D marginalized posteriors from joint constraints from collections of $\TT$ (dark grey; SPT-SZ, \planck{}, ACT DR4 $\ell > 1800$), $\TE$ (red; SPT-3G 2018, \planck{}, ACT DR4), and $\EE$ (blue; SPT-3G 2018, \planck{}, ACT DR4) power spectra for each \lcdm{} parameter vs. $\Hubble$. The solid black contours show constraints from the combination of $\TT$, $\TE$, and $\EE$ spectra from SPT-3G 2018, \planck{}, and ACT DR4. The light grey band indicates the $2\,\sigma$ interval of the distance-ladder measurement of $\Hubble$ by \citetalias{riess20}. Despite the raised expansion rate inferred from each individual $\EE$ spectrum, the joint result is consistent with the $\TT$ and $\TE$ data and remains in $2.2\,\sigma$ tension with the low-redshift measurement of $\Hubble$.
The low acoustic scale value inferred from the $\EE$ spectra is driven by the \planck{} data (see Figure 5 of \citet{planck18-6}).
Contours indicate the 68\% and 95\% probability regions.
}
\label{fig:H0_EE}
\end{figure*}
\begin{table*}[ht!]
\def\arraystretch{1.5}
\small
\setlength{\tabcolsep}{10pt}
\centering
\begin{tabular}{l c c}
\hline
 Spectra & Datasets & $\Hubble\,[\mathrm{km\,s^{-1}\,Mpc^{-1}}]$\\
 \hline
\TT{} & SPT-SZ + \planck{} + ACT DR4 ($\ell > 1800$) & $68.85 \pm 0.97$\\
\TE{} & SPT-3G 2018 + \planck{} + ACT DR4 & $67.95 \pm 0.94$\\
\EE{} & SPT-3G 2018 + \planck{} + ACT DR4 & $69.2 \pm 1.2$\\
$\TT+\TE+\EE$ & SPT-3G 2018 + \planck{} + ACT DR4 & $67.49 \pm 0.53$\\
\hline
\end{tabular}
\caption{
We find consistent constraints on the Hubble constant \Hubble{} for the three spectra, \TT{}, \TE, and \EE, from combinations of SPT, \planck{}, and ACT DR4 datasets.
}
\label{tab:hubble}
\end{table*}

We now turn our attention to the observation made by \citetalias{dutcher21} that current $\EE$ power spectrum measurements are consistent with comparatively high values of $\Hubble$. 
Fits to the $\EE$ power spectra from SPT-3G 2018, SPTpol, \planck{}, and ACT DR4 yield $\Hubble=76.4 \pm 4.1,\; 73.4 \pm 3.3,\; 69.9 \pm 2.7$, and $71.8 \pm 4.4\,\mathrm{km\,s^{-1}\,Mpc^{-1}}$, respectively \citep[\citetalias{dutcher21}]{henning18, aiola20, planck18-5}. 
These values are all within $\leq 1.1\,\sigma$ of the distance-ladder measurement of $\Hubble$ by \citetalias{riess20}. 
As stated by \citetalias{dutcher21}, this inconsistency between cosmological constraints derived from temperature and polarization data might hint at new physics to resolve the Hubble tension. 

Although an interesting lead, the current evidence for such an inconsistency in individual experiments is low (see \citetalias{dutcher21} \S 7, \citet{planck18-6} \S 3, \citet{choi20} \S 12). 
To increase the statistical weight, we combine the measured bandpowers from recent experiments at the likelihood level and present constraints based only on the $\TT$, $\TE$, or $\EE$ spectra. 
For the $\TT$ results we use SPT-SZ, \planck{}, and ACT DR4 data, with the ACT DR4 spectrum limited to the multipole range $\ell > 1800$ as recommended by \citet{aiola20} in order to avoid correlations with the \planck{} data. 
For the $\TE$ and $\EE$ spectra, we combine the SPT-3G 2018, \planck{}, and ACT DR4 data. 
The parameter posteriors for the three sets of spectra are plotted in Figure \ref{fig:H0_EE} and tabulated in Table \ref{tab:hubble}. 
The joint constraints on the expansion rate for the three cases are $\Hubble = 68.85 \pm 0.97\,\mathrm{km\,s^{-1}\,Mpc^{-1}}$ for \TT-only, $\Hubble = 67.95 \pm 0.94\,\mathrm{km\,s^{-1}\,Mpc^{-1}}$ for \TE-only, and $\Hubble = 69.2 \pm 1.2\,\mathrm{km\,s^{-1}\,Mpc^{-1}}$ for \EE-only.
There is no significant shift towards higher expansion rates in the polarization data.
We note that the result from the combined $\EE$ data is lower than the value inferred from each individual dataset.
As discussed by \citet{addison21} and shown by Figure 1 of that work, this is because the ground-based experiments are most consistent with the lower end of the \planck{} $\Hubble$ parameter ellipses.
We conclude that the temperature and polarization constraints paint a consistent picture of a low expansion rate, and do not suggest possible explanations for the gap between the Cepheid and supernova distance-ladder measurements of \citetalias{riess20} and CMB data.




In the late stages of completing this work, \citet{addison21} published a similar, though more extensive, analysis investigating the $\Hubble$ constraints produced by combining $\EE$ power spectra of different experiments. 
While \citet{addison21} use the SPTpol 500d bandpowers, their results are fairly similar to ours.
\citet{addison21} report a combined constraint on $\Hubble$ of $68.7 \pm 1.3\,\mathrm{km\,s^{-1}\,Mpc^{-1}}$ which is consistent with our result of $69.2 \pm 1.2\,\mathrm{km\,s^{-1}\,Mpc^{-1}}$.
Note that the results are not independent, as they use the same data from \planck{} and ACT DR4.  
Moreover, the SPTpol and SPT-3G 2018 datasets produce similar cosmological constraints by themselves as pointed out by \citetalias{dutcher21}, which is partly due to the shared sky area between the two surveys.

We combine the SPT-3G 2018, \planck{}, and ACT DR4 temperature and polarization spectra to obtain the most precise constraint of $\Hubble$ from CMB power spectra to date.\footnote{We exclude SPT-SZ and SPTpol from this comparison due to the shared survey area with SPT-3G.}
We report $\Hubble = 67.49 \pm 0.53\,\mathrm{km\,s^{-1}\,Mpc^{-1}}$. 
This result is $4.1\,\sigma$ lower than the low-redshift measurement of $\Hubble = 73.2 \pm 1.3\,\mathrm{km\,s^{-1}\,Mpc^{-1}}$ by \citetalias{riess20}; the Hubble tension remains. 



\section{Conclusion}
\label{sec:conclusion}

In this work, we have presented constraints on cosmological models beyond \lcdm{} using the SPT-3G 2018 power spectra, paying attention to the results in the context of the Hubble tension. 
The multi-frequency $\EE$ and $\TE$ bandpowers from SPT-3G provide a high-precision measurement of the CMB at intermediate and small angular scales. 
As such, the bandpowers allow us to place tight constraints on physics beyond the standard model.
We look for evidence of models with additional (or fewer) light and free-streaming degrees of freedom, or with non-standard BBN helium production.
Introducing $\Neff$ as a free parameter, we determine $\Neff = 3.70 \pm 0.70$ from SPT-3G 2018 data, which is consistent with the standard model prediction of 3.044 at $0.9\,\sigma$.
Instead varying $\Yp$, we find $\Yp = 0.225 \pm 0.052$, which agrees well with the BBN prediction of 0.2454.
Varying the two parameters simultaneously yields $\Neff = 5.1 \pm 1.2$ and $\Yp = 0.151 \pm 0.060$.
Both values are within $2\,\sigma$ of their \lcdm{} values. 
When adding the SPT-3G data to \planck{}, the constraints tighten to $\Neff = 1.3 \pm 0.3$ and $\Yp = 0.230 \pm 0.017$. 
For the \lcdm{}+$\Neff$ model, the SPT-3G data tighten the \planck{}-only constraints on $\Neff$ and $\Hubble$ by $11\%$.
We see no significant evidence for new light relics or inconsistencies with BBN.

 

We also look at the implications of the SPT-3G 2018 data for the sum of the neutrino masses.
Joint constraints from SPT-3G 2018 and BAO data limit the sum of neutrino masses to $\mnu < 0.30\,\mathrm{eV}$ at 95\% confidence. 
Adding the \planck{} power spectrum data reduces the 95\% CL limit to $\mnu < 0.13\,\mathrm{eV}$. 


We explore the possibility of an additional sterile neutrino, while assuming minimal masses in the normal hierarchy for the three known neutrino species.
From the SPT-3G 2018 data alone we derive a 95\% CL upper limit on the effective mass of $\meff < 1.5\,\mathrm{eV}$ and on the increase to the effective number of neutrino species of $\Delta\Neff < 1.8$. 
Adding BAO data significantly tightens these constraints to $\Delta\Neff < 1.6$ and $\meff < 0.50\,\mathrm{eV}$.

The SPT-3G 2018 dataset is consistent with a flat universe.
We find $\curv = 0.001^{+0.018}_{-0.019}$, which is comparable to the precision of \planck{} data. Adding \planck{} and BAO data refines the constraint by an order of magnitude to $\curv = 0.0009 \pm 0.0018$.

Varying $\Neff$ or $\curv$ allows for higher values of $\Hubble$ with the SPT-3G 2018 data. 
In the first case, the higher values of $\Hubble$ are connected to the slight preference for higher values of $\Neff$ as well as increased uncertainties compared to \lcdm{} constraints. 
The increase in uncertainty is the main effect in the curvature case, where the uncertainty on $\Hubble$ is increased by a factor of 5.3. 
In both cases, the higher values of $\Hubble$ are disfavored by the addition of \planck{} or BAO data.

We find that adding SPT-3G 2018 to \planck{} data reduces the square root of the determinants of the parameter covariance matrices by factors of $1.3 - 2.0$ across the cosmological models considered, signaling a substantial reduction in the allowed parameter volume.

We update the recent work of \citet{addison21}, and combine SPT-3G 2018, \planck{}, and ACT DR4 at the likelihood level and report joint constraints on $\Hubble$ using only the $\EE$ spectra. We find $\Hubble = 69.19 \pm 1.2\,\mathrm{km\,s^{-1}\,Mpc^{-1}}$, which is $2.2\,\sigma$ lower than the distance-ladder measurement of \citetalias{riess20}. We evaluate the significance of the Hubble tension by combining all spectra of the aforementioned datasets to produce the constraint on $\Hubble$ from CMB power spectra to date: $\Hubble = 67.49 \pm 0.53\,\mathrm{km\,s^{-1}\,Mpc^{-1}}$. This value is in $4.1\,\sigma$ tension with the most precise distance-ladder measurement \citepalias{riess20}.


While the SPT-3G 2018 dataset provides a detailed view of the small-scale CMB polarization anisotropy, the data were obtained during a four-month period of the SPT-3G survey, during which approximately half of the detectors were inoperable.
The SPT-3G survey is planned to continue through at least 2023, with existing maps from the combined 2019 and 2020 observing seasons already having $\sim3.5\times$ lower noise than the maps used in this analysis.
The bandpowers from the full SPT-3G survey will significantly improve measurements of the damping tail of the $\TE$ and $\EE$ spectra, enabling tight constraints on physics beyond the \lcdm{} model. 

\acknowledgments
We thank Brian Fields for useful discussions on cosmological models modifying the primordial helium abundance and effective number of neutrino species.
The South Pole Telescope program is supported by the National Science Foundation (NSF) through grants PLR-1248097 and OPP-1852617.
Partial support is also provided by the NSF Physics Frontier Center grant PHY-1125897 to the Kavli Institute of Cosmological Physics at the University of Chicago, the Kavli Foundation, and the Gordon and Betty Moore Foundation through grant GBMF\#947 to the University of Chicago.
Argonne National Laboratory's work was supported by the U.S. Department of Energy, Office of High Energy Physics, under contract DE-AC02-06CH11357.
Work at Fermi National Accelerator Laboratory, a DOE-OS, HEP User Facility managed by the Fermi Research Alliance, LLC, was supported under Contract No. DE-AC02-07CH11359.
The Cardiff authors acknowledge support from the UK Science and Technologies Facilities Council (STFC).
The CU Boulder group acknowledges support from NSF AST-0956135.  
The IAP authors acknowledge support from the Centre National d'\'{E}tudes Spatiales (CNES).
JV acknowledges support from the Sloan Foundation.
The Melbourne authors acknowledge support from the University of Melbourne and an Australian Research Council Future Fellowship (FT150100074). 
The McGill authors acknowledge funding from the Natural Sciences and Engineering Research Council of Canada, Canadian Institute for Advanced Research, and the Fonds de recherche du Qu\'ebec Nature et technologies.
NWH acknowledges support from NSF CAREER grant AST-0956135.
The UCLA and MSU authors acknowledge support from NSF AST-1716965 and CSSI-1835865.
This research was done using resources provided by the Open Science Grid \citep{pordes07, sfiligoi09}, which is supported by the National Science Foundation award 1148698, and the U.S. Department of Energy's Office of Science.
This research used resources of the National Energy Research Scientific Computing Center (NERSC), a U.S. Department of Energy Office of Science User Facility operated under Contract No. DE-AC02-05CH11231.
Some of the results in this paper have been derived using the healpy and HEALPix packages.
The data analysis pipeline also uses the scientific python stack \citep{hunter07, jones01, vanDerWalt11}.
\vspace*{-3mm}

\begin{appendices}
\section*{Appendix}
\subsection{Lensing Convergence on the SPT-3G Survey Field}
\label{app:kappa}

The matter density field between us and recombination lenses the CMB and changes the observed power spectrum. One non-trivial consequence of this for surveys that do not cover a large fraction of the sky is super-sample lensing, i.e. the distortion of the CMB caused by matter-fluctuation modes with wavelengths larger than the survey field. 
This effect can be accounted for by adding a term to the covariance matrix or by marginalizing over the mean convergence across the survey field, $\bar{\kappa}$ \citep{manzotti14}.
While both yield the same results, we have chosen the latter approach in this work because it has the advantage of returning information on the local matter density across the survey field. 
As such, introducing $\bar{\kappa}$ as a variable in the MCMC chains can help us better understand the data and provide context when comparing the SPT-3G results to those of other experiments. 

We account for super-sample lensing in our likelihood analysis by modifying the model spectrum, $ C_\ell(p)$, based on a number of parameters $\mathbf{p}$ to $\hat{C}_\ell(\mathbf{p},\bar{\kappa})$ via
\begin{equation}
\hat{C}_\ell(\mathbf{p},\bar{\kappa}) = C_\ell(\mathbf{p}) + \frac{\partial\ell^2 C_\ell(p)}{\partial \ln{\ell}} \frac{\bar{\kappa}}{\ell^2}.
\end{equation}
Note that the definition of $\bar{\kappa}$ is of opposite sign to \citetalias{dutcher21}, matching \citet{motloch19}.
All cosmological constraints presented in this work have been derived using a Gaussian prior centered on zero with width $4.5 \times 10^{-4}$ as shown in Table \ref{tab:priors_table}. The prior width is based on the geometry of the survey field \citep{manzotti14}.

\begin{figure}[t!]
\includegraphics[width=8.6cm]{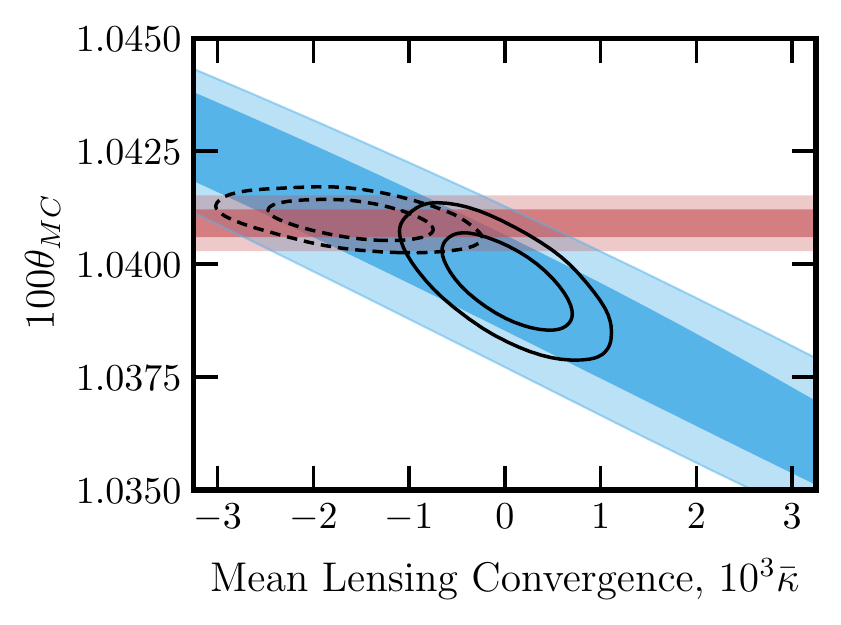}
\centering
\caption{
Constraints in the $\bar{\kappa}$ vs. $\theta_{MC}$ plane from SPT-3G 2018 without (blue contour) and with (solid black line) a prior on the mean convergence. Contours indicate the 68\% and 95\% probability regions. Using SPT-3G 2018 data alone the two parameters are degenerate with one another, unless a prior is placed on the mean convergence. The red band indicates the $2\sigma$ range of the latest \planck{} value for $\theta_{MC}$. From the joint constraints from SPT-3G 2018 and \planck{} without a prior on $\bar{\kappa}$ (dashed black lines) we infer $\bar{\kappa}<0$ at $2.9\sigma$.}
\label{fig:kappa_theta}
\end{figure}

Due to the limited sky fraction observed by SPT-3G, $\bar{\kappa}$ is degenerate with $\theta_{MC}$ as can be seen in Figure \ref{fig:kappa_theta}. 
This degeneracy was already noted by \citet{motloch19} using the example of the SPTpol 500d dataset, which for this purpose is similar to the SPT-3G dataset.
The $\bar{\kappa}-\theta$ degeneracy can be broken by imposing a prior on $\bar{\kappa}$ centered on zero with a width that depends on the field area, as we have done throughout this work.

\citet{motloch19} demonstrate that the $\bar{\kappa}-\theta$ degeneracy can be broken without resorting to a prior on $\bar{\kappa}$ through the inclusion of \planck{} data, which due to its large sky coverage is insensitive to super-sample lensing. 
Combining SPT-3G 2018 and \planck{} data yields an estimate of the mean convergence on the SPT-3G survey field,
\begin{equation}
10^3 \bar{\kappa}_{\rm SPT-3G} = -1.60 \pm 0.56.
\end{equation}
This inferred $\bar{\kappa}$ is $2.9\,\sigma$ away from zero and would imply that the SPT-3G footprint coincides with a local underdensity.
If we include the expected \lcdm{} cosmic variance across this field size (0.45 as mentioned above), this becomes a 2.2\,$\sigma$ event.

We run SPT-3G-only chains imposing this result as a prior on $\bar{\kappa}$ instead of the zero-centred prior used throughout this work (see Table \ref{tab:priors_table}).
As expected, the $\theta_{MC}$ constraint shifts high to $1.04126 \pm 0.00078$, which is close to the \planck{} result ($1.04090 \pm 0.00031$).
The central values of other \lcdm{} parameters only shift slightly ($\lesssim 0.1\,\sigma$). The inferred $\Hubble$ changes from $68.8 \pm 1.5\,\kmsmpc$ to $69.2 \pm 1.5\,\kmsmpc$.

\subsection{Parameter Tables}
\label{app:tables}
\vspace*{-2mm}
We present the full parameter constraints from SPT-3G 2018 alone and in combination with BAO and \planck{} data on \lcdm{} extensions in the following tables. We show results for \lcdm{}+$\Neff$, \lcdm{}+$\Yp$, and \lcdm{}+$\Neff$+$\Yp$ in Table \ref{tab:neff_param_table}. We show constraints on \lcdm{}+$\mnu$ and \lcdm{}+$\meff$ in tables \ref{tab:mnu_param_table} and \ref{tab:meff_param_table}, respectively. Constraints on \lcdm{}+$\curv$ are given in Table \ref{tab:curv_param_table}.
\begin{table*}[ht!]
\def\arraystretch{1.2}
\footnotesize
\setlength{\tabcolsep}{3pt}
\centering
\begin{tabular}{c @{\hskip 10pt} D{+}{\,\pm\,}{7.7} D{+}{\,\pm\,}{7.7} D{+}{\,\pm\,}{7.7} D{+}{\,\pm\,}{7.7} D{+}{\,\pm\,}{7.7} D{+}{\,\pm\,}{7.7} }
& \multicolumn{2}{c}{\text{$N_{\rm eff}$}}
& \multicolumn{2}{c}{\text{$Y_P$}}
& \multicolumn{2}{c}{\text{$N_{\rm eff} + Y_P$}} \\
 \hline \vspace{-8pt} \\
& \multicolumn{1}{c}{SPT-3G 2018} & \multicolumn{1}{c}{\makecell{SPT-3G 2018\\ + \planck{}}} & \multicolumn{1}{c}{SPT-3G 2018} & \multicolumn{1}{c}{\makecell{SPT-3G 2018\\ + \planck{}}} & \multicolumn{1}{c}{SPT-3G 2018} & \multicolumn{1}{c}{\makecell{SPT-3G 2018\\ + \planck{}}}\\
 \hline
\multicolumn{4}{l}{Free}\\
$\Omega_b h^2$ &
0.02275+0.00048 &
0.02232+0.00020 &
0.02231+0.00050 &
0.02229+0.00019 &
0.02256+0.00049 &
0.02230+0.00020 \\
$\Omega_c h^2$ &
0.1232+0.0097 &
0.1183+0.0027 &
0.1152+0.0037 &
0.1197+0.0013 &
0.141+0.016 &
0.1210+0.0045 \\
$100\theta_{MC}$ &
1.03913+0.00089 &
1.04086+0.00039 &
1.0390+0.0018 &
1.04034+0.00051 &
1.0345+0.0027 &
1.0400+0.0011 \\
$10^9 A_s e^{-2\tau}$ &
1.828+0.041 &
1.873+0.016 &
1.824+0.038 &
1.876+0.012 &
1.866+0.046 &
1.879+0.018 \\
$n_s$ &
1.038+0.046 &
0.9629+0.0079 &
0.984+0.044 &
0.9615+0.0068 &
1.019+0.046 &
0.9627+0.0079 \\
$N_{eff}$ &
3.70+0.70 &
2.95+0.17 &
\multicolumn{1}{D{+}{\,-\,}{7.7}}{+} &
\multicolumn{1}{D{+}{\,-\,}{7.7}}{+} &
5.1+1.2 &
3.13+0.30 \\
$Y_P$ &
\multicolumn{1}{D{+}{\,-\,}{7.7}}{+} &
\multicolumn{1}{D{+}{\,-\,}{7.7}}{+} &
0.225+0.052 &
0.234+0.012 &
0.151+0.060 &
0.230+0.017 \\
\hline
\multicolumn{7}{l}{Derived}\\
$H_0$ &
73.5+5.2 &
66.8+1.3 &
68.4+1.7 &
67.20+0.63 &
80.4+7.2 &
67.7+1.8 \\
$\Omega_\Lambda$ &
0.726+0.028 &
0.6833+0.0095 &
0.704+0.022 &
0.6839+0.0083 &
0.743+0.027 &
0.6854+0.0099 \\
$\sigma_8$ &
0.812+0.030 &
0.804+0.010 &
0.786+0.020 &
0.8058+0.0077 &
0.829+0.032 &
0.808+0.012 \\
$S_8$ &
0.774+0.042 &
0.826+0.015 &
0.780+0.041 &
0.827+0.015 &
0.765+0.042 &
0.827+0.014 \\
${\rm{Age}}/{\rm{Gyr}}$ &
13.22+0.63 &
13.90+0.18 &
13.84+0.10 &
13.822+0.034 &
12.32+0.80 &
13.75+0.27 \\
\hline
\end{tabular}
\caption[
$\Lambda$CDM parameter constraints.
]{
Constraints on $\Lambda$CDM model extensions $\Neff$, $\Yp$, and $\Neff+\Yp$ from SPT-3G 2018 alone and jointly with \planck{}.
}
\label{tab:neff_param_table}
\end{table*}
\begin{table*}[hb!]
\def\arraystretch{1.2}
\footnotesize
\setlength{\tabcolsep}{3pt}
\centering
\begin{tabular}{c @{\hskip 10pt} D{+}{\,\pm\,}{7.7} D{+}{\,\pm\,}{7.7} D{+}{\,\pm\,}{7.7} D{+}{\,\pm\,}{7.7} }
& \multicolumn{4}{c}{\text{$\mnu$}} \\
 \hline \vspace{-8pt} \\
& \multicolumn{1}{c}{SPT-3G 2018} & \multicolumn{1}{c}{\makecell{SPT-3G 2018\\ + \planck{}}} & \multicolumn{1}{c}{\makecell{SPT-3G 2018\\ + BAO}} & \multicolumn{1}{c}{\makecell{SPT-3G 2018\\ + \planck{} + BAO}}\\
 \hline
\multicolumn{5}{l}{Free}\\
$\Omega_b h^2$ &
0.02239+0.00033 &
0.02239+0.00014 &
0.02244+0.00032 &
0.02246+0.00012 \\
$\Omega_c h^2$ &
0.1179+0.0042 &
0.1197+0.0013 &
0.1152+0.0019 &
0.11885+0.00099 \\
$100\theta_{MC}$ &
1.03907+0.00082 &
1.04070+0.00029 &
1.03956+0.00066 &
1.04082+0.00027 \\
$10^9 A_s e^{-2\tau}$ &
1.838+0.041 &
1.880+0.011 &
1.824+0.036 &
1.877+0.010 \\
$n_s$ &
0.980+0.026 &
0.9662+0.0043 &
0.997+0.018 &
0.9682+0.0037 \\
$\Sigma m_\nu$ &
 \multicolumn{1}{D{+}{\,<\,}{7.7}}{+2.0} &
 \multicolumn{1}{D{+}{\,<\,}{7.7}}{+0.29} &
 \multicolumn{1}{D{+}{\,<\,}{7.7}}{+0.30} &
 \multicolumn{1}{D{+}{\,<\,}{7.7}}{+0.13} \\
\hline
\multicolumn{5}{l}{Derived}\\
$H_0$ &
62.7+5.3 &
67.1+1.1 &
68.02+0.70 &
67.92+0.52 \\
$\Omega_\Lambda$ &
0.61+0.11 &
0.681+0.015 &
0.6991+0.0087 &
0.6924+0.0067 \\
$\sigma_8$ &
0.686+0.089 &
0.801+0.021 &
0.774+0.025 &
0.810+0.011 \\
$S_8$ &
0.764+0.045 &
0.825+0.016 &
0.775+0.027 &
0.820+0.013 \\
${\rm{Age}}/{\rm{Gyr}}$ &
14.11+0.27 &
13.820+0.059 &
13.847+0.052 &
13.779+0.027 \\
\hline
\end{tabular}
\caption[
$\Lambda$CDM parameter constraints.
]{
Combined constraints on $\Lambda$CDM model extension $\mnu$ from the SPT-3G 2018, \planck{}, and BAO datasets.
}
\label{tab:mnu_param_table}
\end{table*}
\begin{table*}[ht!]
\def\arraystretch{1.2}
\footnotesize
\setlength{\tabcolsep}{3pt}
\centering
\begin{tabular}{c @{\hskip 10pt} D{+}{\,\pm\,}{7.7} D{+}{\,\pm\,}{7.7} D{+}{\,\pm\,}{7.7} D{+}{\,\pm\,}{7.7} }
& \multicolumn{4}{c}{\text{$\meff$}} \\
 \hline \vspace{-8pt} \\
& \multicolumn{1}{c}{SPT-3G 2018} & \multicolumn{1}{c}{\makecell{SPT-3G 2018\\ + \planck{}}} & \multicolumn{1}{c}{\makecell{SPT-3G 2018\\ + BAO}} & \multicolumn{1}{c}{\makecell{SPT-3G 2018\\ + \planck{} + BAO}}\\
 \hline
\multicolumn{5}{l}{Free}\\
$\Omega_b h^2$ &
0.02284+0.00042 &
0.02248+0.00014 &
0.02281+0.00039 &
0.02256+0.00013 \\
$\Omega_c h^2$ &
0.1278+0.0079 &
0.1210+0.0019 &
0.1269+0.0077 &
0.1201+0.0018 \\
$100\theta_{MC}$ &
1.03858+0.00082 &
1.04052+0.00032 &
1.03877+0.00078 &
1.04066+0.00031 \\
$10^9 A_s e^{-2\tau}$ &
1.841+0.042 &
1.888+0.013 &
1.844+0.037 &
1.883+0.012 \\
$n_s$ &
1.042+0.036 &
0.9690+0.0053 &
1.038+0.031 &
0.9725+0.0050 \\
$\Delta N_{eff}$ &
 \multicolumn{1}{D{+}{\,<\,}{7.7}}{+1.8} &
 \multicolumn{1}{D{+}{\,<\,}{7.7}}{+0.30} &
 \multicolumn{1}{D{+}{\,<\,}{7.7}}{+1.6} &
 \multicolumn{1}{D{+}{\,<\,}{7.7}}{+0.30} \\
$m_{\nu,{\rm{sterile}}}^{\rm{eff}}$ &
 \multicolumn{1}{D{+}{\,<\,}{7.7}}{+1.5} &
 \multicolumn{1}{D{+}{\,<\,}{7.7}}{+0.44} &
 \multicolumn{1}{D{+}{\,<\,}{7.7}}{+0.50} &
 \multicolumn{1}{D{+}{\,<\,}{7.7}}{+0.20} \\
\hline
\multicolumn{5}{l}{Derived}\\
$H_0$ &
71.0+4.4 &
67.47+0.81 &
71.6+2.2 &
68.30+0.70 \\
$\Omega_\Lambda$ &
0.686+0.044 &
0.680+0.011 &
0.7020+0.0086 &
0.6911+0.0065 \\
$\sigma_8$ &
0.741+0.063 &
0.787+0.021 &
0.777+0.030 &
0.798+0.013 \\
$S_8$ &
0.753+0.047 &
0.813+0.018 &
0.774+0.031 &
0.810+0.014 \\
${\rm{Age}}/{\rm{Gyr}}$ &
13.16+0.41 &
13.713+0.073 &
13.20+0.37 &
13.687+0.085 \\
\hline
\end{tabular}
\caption[
$\Lambda$CDM parameter constraints.
]{
Combined constraints on $\Lambda$CDM model extension $\meff$ from the SPT-3G 2018, \planck{}, and BAO datasets.
}
\label{tab:meff_param_table}
\end{table*}
\begin{table*}
\def\arraystretch{1.2}
\footnotesize
\setlength{\tabcolsep}{3pt}
\centering
\begin{tabular}{c @{\hskip 10pt} D{+}{\,\pm\,}{7.7} D{+}{\,\pm\,}{7.7} D{+}{\,\pm\,}{7.7} D{+}{\,\pm\,}{7.7} }
& \multicolumn{4}{c}{\text{$\curv$}} \\
 \hline \vspace{-8pt} \\
& \multicolumn{1}{c}{SPT-3G 2018} & \multicolumn{1}{c}{\makecell{SPT-3G 2018\\ + \planck{}}} & \multicolumn{1}{c}{\makecell{SPT-3G 2018\\ + BAO}} & \multicolumn{1}{c}{\makecell{SPT-3G 2018\\ + \planck{} + BAO}}\\
 \hline
\multicolumn{5}{l}{Free}\\
$\Omega_b h^2$ &
0.02241+0.00033 &
0.02251+0.00015 &
0.02243+0.00033 &
0.02242+0.00014 \\
$\Omega_c h^2$ &
0.1162+0.0055 &
0.1184+0.0014 &
0.1149+0.0038 &
0.1192+0.0013 \\
$100\theta_{MC}$ &
1.03956+0.00081 &
1.04086+0.00030 &
1.03960+0.00073 &
1.04075+0.00028 \\
$10^9 A_s e^{-2\tau}$ &
1.828+0.045 &
1.875+0.011 &
1.822+0.039 &
1.877+0.011 \\
$\Omega_K$ &
\multicolumn{1}{D{a}{}{4.6}}{0.001a^{+0.018}_{-0.019}}&
-0.020+0.011 &
-0.0014+0.0037 &
0.0009+0.0018 \\
\hline
\multicolumn{5}{l}{Derived}\\
$H_0$ &
70.8+8.5 &
60.6+3.4 &
68.11+0.76 &
68.05+0.67 \\
$\Omega_\Lambda$ &
0.710+0.046 &
0.630+0.032 &
0.704+0.011 &
0.6918+0.0059 \\
$\sigma_8$ &
0.794+0.030 &
0.789+0.012 &
0.788+0.017 &
0.8082+0.0077 \\
$S_8$ &
0.772+0.068 &
0.897+0.039 &
0.785+0.027 &
0.818+0.012 \\
${\rm{Age}}/{\rm{Gyr}}$ &
13.65+0.92 &
14.57+0.39 &
13.88+0.16 &
13.751+0.077 \\
\hline
\end{tabular}
\caption[
$\Lambda$CDM parameter constraints.
]{
Combined constraints on $\Lambda$CDM model extension $\curv$ from the SPT-3G 2018, \planck{}, and BAO datasets.
}
\label{tab:curv_param_table}
\end{table*}
\clearpage
\end{appendices}

\bibliography{../../BIBTEX/spt}

\begin{thebibliography}{}
\expandafter\ifx\csname natexlab\endcsname\relax\def\natexlab#1{#1}\fi
\providecommand{\url}[1]{\href{#1}{#1}}
\providecommand{\dodoi}[1]{doi:~\href{http://doi.org/#1}{\nolinkurl{#1}}}
\providecommand{\doeprint}[1]{\href{http://ascl.net/#1}{\nolinkurl{http://ascl.net/#1}}}
\providecommand{\doarXiv}[1]{\href{https://arxiv.org/abs/#1}{\nolinkurl{https://arxiv.org/abs/#1}}}

\bibitem[{{Planck Collaboration} {et~al.}(2020{\natexlab{a}}){Planck
  Collaboration}, {Aghanim}, {Akrami}, {Ashdown}, {Aumont}, {Baccigalupi},
  {Ballardini}, {Banday}, {Barreiro}, {Bartolo}, {Basak}, {Benabed}, {Bernard},
  {Bersanelli}, {Bielewicz}, {Bock}, {Bond}, {Borrill}, {Bouchet}, {Boulanger},
  {Bucher}, {Burigana}, {Butler}, {Calabrese}, {Cardoso}, {Carron},
  {Casaponsa}, {Challinor}, {Chiang}, {Colombo}, {Combet}, {Crill}, {Cuttaia},
  {de Bernardis}, {de Rosa}, {de Zotti}, {Delabrouille}, {Delouis}, {Di
  Valentino}, {Diego}, {Dor{\'e}}, {Douspis}, {Ducout}, {Dupac}, {Dusini},
  {Efstathiou}, {Elsner}, {En{\ss}lin}, {Eriksen}, {Fantaye}, {Fernand
  ez-Cobos}, {Finelli}, {Frailis}, {Fraisse}, {Franceschi}, {Frolov},
  {Galeotta}, {Galli}, {Ganga}, {G{\'e}nova-Santos}, {Gerbino}, {Ghosh},
  {Giraud-H{\'e}raud}, {Gonz{\'a}lez-Nuevo}, {G{\'o}rski}, {Gratton},
  {Gruppuso}, {Gudmundsson}, {Hamann}, {Handley}, {Hansen}, {Herranz}, {Hivon},
  {Huang}, {Jaffe}, {Jones}, {Keih{\"a}nen}, {Keskitalo}, {Kiiveri}, {Kim},
  {Kisner}, {Krachmalnicoff}, {Kunz}, {Kurki-Suonio}, {Lagache}, {Lamarre},
  {Lasenby}, {Lattanzi}, {Lawrence}, {Le Jeune}, {Levrier}, {Lewis}, {Liguori},
  {Lilje}, {Lilley}, {Lindholm}, {L{\'o}pez-Caniego}, {Lubin}, {Ma},
  {Mac{\'\i}as-P{\'e}rez}, {Maggio}, {Maino}, {Mandolesi}, {Mangilli},
  {Marcos-Caballero}, {Maris}, {Martin}, {Mart{\'\i}nez-Gonz{\'a}lez},
  {Matarrese}, {Mauri}, {McEwen}, {Meinhold}, {Melchiorri}, {Mennella},
  {Migliaccio}, {Millea}, {Miville-Desch{\^e}nes}, {Molinari}, {Moneti},
  {Montier}, {Morgante}, {Moss}, {Natoli}, {N{\o}rgaard-Nielsen}, {Pagano},
  {Paoletti}, {Partridge}, {Patanchon}, {Peiris}, {Perrotta}, {Pettorino},
  {Piacentini}, {Polenta}, {Puget}, {Rachen}, {Reinecke}, {Remazeilles},
  {Renzi}, {Rocha}, {Rosset}, {Roudier}, {Rubi{\~n}o-Mart{\'\i}n},
  {Ruiz-Granados}, {Salvati}, {Sandri}, {Savelainen}, {Scott}, {Shellard},
  {Sirignano}, {Sirri}, {Spencer}, {Sunyaev}, {Suur-Uski}, {Tauber},
  {Tavagnacco}, {Tenti}, {Toffolatti}, {Tomasi}, {Trombetti}, {Valiviita}, {Van
  Tent}, {Vielva}, {Villa}, {Vittorio}, {Wandelt}, {Wehus}, {Zacchei}, \&
  {Zonca}}]{planck18-5}
{Planck Collaboration}, {Aghanim}, N., {Akrami}, Y., {et~al.}
\newblock {Planck 2018 results. V. CMB power spectra and likelihoods}.
  2020{\natexlab{a}}, \aap, 641, A5, \dodoi{10.1051/0004-6361/201936386}

\bibitem[{{Aiola} {et~al.}(2020){Aiola}, {Calabrese}, {Maurin}, {Naess},
  {Schmitt}, {Abitbol}, {Addison}, {Ade}, {Alonso}, {Amiri}, {Amodeo},
  {Angile}, {Austermann}, {Baildon}, {Battaglia}, {Beall}, {Bean}, {Becker},
  {Bond}, {Bruno}, {Calafut}, {Campusano}, {Carrero}, {Chesmore}, {Cho.},
  {Choi}, {Clark}, {Cothard}, {Crichton}, {Crowley}, {Darwish}, {Datta},
  {Denison}, {Devlin}, {Duell}, {Duff}, {Duivenvoorden}, {Dunkley},
  {D{\"u}nner}, {Essinger-Hileman}, {Fankhanel}, {Ferraro}, {Fox}, {Fuzia},
  {Gallardo}, {Gluscevic}, {Golec}, {Grace}, {Gralla}, {Guan}, {Hall},
  {Halpern}, {Han}, {Hargrave}, {Hasselfield}, {Helton}, {Henderson},
  {Hensley}, {Hill}, {Hilton}, {Hilton}, {Hincks}, {Hlo{\v{z}}ek}, {Ho},
  {Hubmayr}, {Huffenberger}, {Hughes}, {Infante}, {Irwin}, {Jackson}, {Klein},
  {Knowles}, {Koopman}, {Kosowsky}, {Lakey}, {Li}, {Li}, {Li}, {Lokken},
  {Louis}, {Lungu}, {MacInnis}, {Madhavacheril}, {Maldonado}, {Mallaby-Kay},
  {Marsden}, {McMahon}, {Menanteau}, {Moodley}, {Morton}, {Namikawa}, {Nati},
  {Newburgh}, {Nibarger}, {Nicola}, {Niemack}, {Nolta}, {Orlowski-Sherer},
  {Page}, {Pappas}, {Partridge}, {Phakathi}, {Prince}, {Puddu}, {Qu}, {Rivera},
  {Robertson}, {Rojas}, {Salatino}, {Schaan}, {Schillaci}, {Sehgal}, {Sherwin},
  {Sierra}, {Sievers}, {Sifon}, {Sikhosana}, {Simon}, {Spergel}, {Staggs},
  {Stevens}, {Storer}, {Sunder}, {Switzer}, {Thorne}, {Thornton}, {Trac},
  {Treu}, {Tucker}, {Vale}, {Van Engelen}, {Van Lanen}, {Vavagiakis},
  {Wagoner}, {Wang}, {Ward}, {Wollack}, {Xu}, {Zago}, \& {Zhu}}]{aiola20}
{Aiola}, S., {Calabrese}, E., {Maurin}, L., {et~al.}
\newblock {The Atacama Cosmology Telescope: DR4 Maps and Cosmological
  Parameters}. 2020, arXiv e-prints, arXiv:2007.07288.
\newblock \doarXiv{2007.07288}

\bibitem[{{Choi} {et~al.}(2020){Choi}, {Hasselfield}, {Ho}, {Koopman}, {Lungu},
  {Abitbol}, {Addison}, {Ade}, {Aiola}, {Alonso}, {Amiri}, {Amodeo}, {Angile},
  {Austermann}, {Baildon}, {Battaglia}, {Beall}, {Bean}, {Becker}, {Bond},
  {Bruno}, {Calabrese}, {Calafut}, {Campusano}, {Carrero}, {Chesmore}, {Cho.},
  {Clark}, {Cothard}, {Crichton}, {Crowley}, {Darwish}, {Datta}, {Denison},
  {Devlin}, {Duell}, {Duff}, {Duivenvoorden}, {Dunkley}, {D{\"u}nner},
  {Essinger-Hileman}, {Fankhanel}, {Ferraro}, {Fox}, {Fuzia}, {Gallardo},
  {Gluscevic}, {Golec}, {Grace}, {Gralla}, {Guan}, {Hall}, {Halpern}, {Han},
  {Hargrave}, {Henderson}, {Hensley}, {Hill}, {Hilton}, {Hilton}, {Hincks},
  {Hlo{\v{z}}ek}, {Hubmayr}, {Huffenberger}, {Hughes}, {Infante}, {Irwin},
  {Jackson}, {Klein}, {Knowles}, {Kosowsky}, {Lakey}, {Li}, {Li}, {Li},
  {Lokken}, {Louis}, {MacInnis}, {Madhavacheril}, {Maldonado}, {Mallaby-Kay},
  {Marsden}, {Maurin}, {McMahon}, {Menanteau}, {Moodley}, {Morton}, {Naess},
  {Namikawa}, {Nati}, {Newburgh}, {Nibarger}, {Nicola}, {Niemack}, {Nolta},
  {Orlowski-Sherer}, {Page}, {Pappas}, {Partridge}, {Phakathi}, {Prince},
  {Puddu}, {Qu}, {Rivera}, {Robertson}, {Rojas}, {Salatino}, {Schaan},
  {Schillaci}, {Schmitt}, {Sehgal}, {Sherwin}, {Sierra}, {Sievers}, {Sifon},
  {Sikhosana}, {Simon}, {Spergel}, {Staggs}, {Stevens}, {Storer}, {Sunder},
  {Switzer}, {Thorne}, {Thornton}, {Trac}, {Treu}, {Tucker}, {Vale}, {Van
  Engelen}, {Van Lanen}, {Vavagiakis}, {Wagoner}, {Wang}, {Ward}, {Wollack},
  {Xu}, {Zago}, \& {Zhu}}]{choi20}
{Choi}, S.~K., {Hasselfield}, M., {Ho}, S.-P.~P., {et~al.}
\newblock {The Atacama Cosmology Telescope: A Measurement of the Cosmic
  Microwave Background Power Spectra at 98 and 150 GHz}. 2020, arXiv e-prints,
  arXiv:2007.07289.
\newblock \doarXiv{2007.07289}

\bibitem[{{Reichardt} {et~al.}(2020){Reichardt}, {Patil}, {Ade}, {Anderson},
  {Austermann}, {Avva}, {Baxter}, {Beall}, {Bender}, {Benson}, {Bianchini},
  {Bleem}, {Carlstrom}, {Chang}, {Chaubal}, {Chiang}, {Chou}, {Citron},
  {Corbett Moran}, {Crawford}, {Crites}, {de Haan}, {Dobbs}, {Everett},
  {Gallicchio}, {George}, {Gilbert}, {Gupta}, {Halverson}, {Harrington},
  {Henning}, {Hilton}, {Holder}, {Holzapfel}, {Hrubes}, {Huang}, {Hubmayr},
  {Irwin}, {Knox}, {Lee}, {Li}, {Lowitz}, {Luong-Van}, {McMahon}, {Mehl},
  {Meyer}, {Millea}, {Mocanu}, {Mohr}, {Montgomery}, {Nadolski}, {Natoli},
  {Nibarger}, {Noble}, {Novosad}, {Omori}, {Padin}, {Pryke}, {Ruhl},
  {Saliwanchik}, {Sayre}, {Schaffer}, {Shirokoff}, {Sievers}, {Smecher},
  {Spieler}, {Staniszewski}, {Stark}, {Tucker}, {Vand erlinde}, {Veach},
  {Vieira}, {Wang}, {Whitehorn}, {Williamson}, {Wu}, \&
  {Yefremenko}}]{reichardt20}
{Reichardt}, C.~L., {Patil}, S., {Ade}, P.~A.~R., {et~al.}
\newblock {An Improved Measurement of the Secondary Cosmic Microwave Background
  Anisotropies from the SPT-SZ + SPTpol Surveys}. 2020, arXiv e-prints,
  arXiv:2002.06197.
\newblock \doarXiv{2002.06197}

\bibitem[{{Planck Collaboration} {et~al.}(2016{\natexlab{a}}){Planck
  Collaboration}, {Aghanim}, {Arnaud}, {Ashdown}, {Aumont}, {Baccigalupi},
  {Banday}, {Barreiro}, {Bartlett}, {Bartolo}, \& et~al.}]{planck15-11}
{Planck Collaboration}, {Aghanim}, N., {Arnaud}, M., {Ashdown}, M., {Aumont},
  J., {Baccigalupi}, C., {Banday}, A.~J., {Barreiro}, R.~B., {Bartlett}, J.~G.,
  {Bartolo}, N., \& et~al.
\newblock {Planck 2015 results. XI. CMB power spectra, likelihoods, and
  robustness of parameters}. 2016{\natexlab{a}}, \aap, 594, A11,
  \dodoi{10.1051/0004-6361/201526926}

\bibitem[{{Planck Collaboration} {et~al.}(2017){Planck Collaboration},
  {Aghanim}, {Akrami}, {Ashdown}, {Aumont}, {Baccigalupi}, {Ballardini},
  {Banday}, {Barreiro}, {Bartolo}, {Basak}, {Benabed}, {Bersanelli},
  {Bielewicz}, {Bonaldi}, {Bonavera}, {Bond}, {Borrill}, {Bouchet}, {Burigana},
  {Calabrese}, {Cardoso}, {Challinor}, {Chiang}, {Colombo}, {Combet}, {Crill},
  {Curto}, {Cuttaia}, {de Bernardis}, {de Rosa}, {de Zotti}, {Delabrouille},
  {Di Valentino}, {Dickinson}, {Diego}, {Dor{\'e}}, {Ducout}, {Dupac},
  {Dusini}, {Efstathiou}, {Elsner}, {En{\ss}lin}, {Eriksen}, {Fantaye},
  {Finelli}, {Forastieri}, {Frailis}, {Franceschi}, {Frolov}, {Galeotta},
  {Galli}, {Ganga}, {G{\'e}nova-Santos}, {Gerbino}, {Gonz{\'a}lez-Nuevo},
  {G{\'o}rski}, {Gratton}, {Gruppuso}, {Gudmundsson}, {Herranz}, {Hivon},
  {Huang}, {Jaffe}, {Jones}, {Keih{\"a}nen}, {Keskitalo}, {Kiiveri}, {Kim},
  {Kisner}, {Knox}, {Krachmalnicoff}, {Kunz}, {Kurki-Suonio}, {Lagache},
  {Lamarre}, {Lasenby}, {Lattanzi}, {Lawrence}, {Le Jeune}, {Levrier}, {Lewis},
  {Liguori}, {Lilje}, {Lilley}, {Lindholm}, {L{\'o}pez-Caniego}, {Lubin}, {Ma},
  {Mac{\'{\i}}as-P{\'e}rez}, {Maggio}, {Maino}, {Mandolesi}, {Mangilli},
  {Maris}, {Martin}, {Mart{\'{\i}}nez-Gonz{\'a}lez}, {Matarrese}, {Mauri},
  {McEwen}, {Meinhold}, {Mennella}, {Migliaccio}, {Millea},
  {Miville-Desch{\^e}nes}, {Molinari}, {Moneti}, {Montier}, {Morgante}, {Moss},
  {Narimani}, {Natoli}, {Oxborrow}, {Pagano}, {Paoletti}, {Partridge},
  {Patanchon}, {Patrizii}, {Pettorino}, {Piacentini}, {Polastri}, {Polenta},
  {Puget}, {Rachen}, {Racine}, {Reinecke}, {Remazeilles}, {Renzi}, {Rocha},
  {Rossetti}, {Roudier}, {Rubi{\~n}o-Mart{\'{\i}}n}, {Ruiz-Granados},
  {Salvati}, {Sandri}, {Savelainen}, {Scott}, {Sirignano}, {Sirri}, {Stanco},
  {Suur-Uski}, {Tauber}, {Tavagnacco}, {Tenti}, {Toffolatti}, {Tomasi},
  {Tristram}, {Trombetti}, {Valiviita}, {Van Tent}, {Vielva}, {Villa},
  {Vittorio}, {Wandelt}, {Wehus}, {White}, {Zacchei}, \& {Zonca}}]{planck16-51}
{Planck Collaboration}, {Aghanim}, N., {Akrami}, Y., {et~al.}
\newblock {Planck intermediate results. LI. Features in the cosmic microwave
  background temperature power spectrum and shifts in cosmological parameters}.
  2017, \aap, 607, A95, \dodoi{10.1051/0004-6361/201629504}

\bibitem[{{Henning} {et~al.}(2018){Henning}, {Sayre}, {Reichardt}, {Ade},
  {Anderson}, {Austermann}, {Beall}, {Bender}, {Benson}, {Bleem}, {Carlstrom},
  {Chang}, {Chiang}, {Cho}, {Citron}, {Corbett Moran}, {Crawford}, {Crites},
  {de Haan}, {Dobbs}, {Everett}, {Gallicchio}, {George}, {Gilbert},
  {Halverson}, {Harrington}, {Hilton}, {Holder}, {Holzapfel}, {Hoover}, {Hou},
  {Hrubes}, {Huang}, {Hubmayr}, {Irwin}, {Keisler}, {Knox}, {Lee}, {Leitch},
  {Li}, {Lowitz}, {Manzotti}, {McMahon}, {Meyer}, {Mocanu}, {Montgomery},
  {Nadolski}, {Natoli}, {Nibarger}, {Novosad}, {Padin}, {Pryke}, {Ruhl},
  {Saliwanchik}, {Schaffer}, {Sievers}, {Smecher}, {Stark}, {Story}, {Tucker},
  {Vanderlinde}, {Veach}, {Vieira}, {Wang}, {Whitehorn}, {Wu}, \&
  {Yefremenko}}]{henning18}
{Henning}, J.~W., {Sayre}, J.~T., {Reichardt}, C.~L., {et~al.}
\newblock {Measurements of the Temperature and E-mode Polarization of the CMB
  from 500 Square Degrees of SPTpol Data}. 2018, \apj, 852, 97,
  \dodoi{10.3847/1538-4357/aa9ff4}

\bibitem[{{Aylor} {et~al.}(2017){Aylor}, {Hou}, {Knox}, {Story}, {Benson},
  {Bleem}, {Carlstrom}, {Chang}, {Cho}, {Chown}, {Crawford}, {Crites}, {de
  Haan}, {Dobbs}, {Everett}, {George}, {Halverson}, {Harrington}, {Holder},
  {Holzapfel}, {Hrubes}, {Keisler}, {Lee}, {Leitch}, {Luong-Van}, {Marrone},
  {McMahon}, {Meyer}, {Millea}, {Mocanu}, {Mohr}, {Natoli}, {Omori}, {Padin},
  {Pryke}, {Reichardt}, {Ruhl}, {Sayre}, {Schaffer}, {Shirokoff},
  {Staniszewski}, {Stark}, {Vanderlinde}, {Vieira}, \& {Williamson}}]{aylor17}
{Aylor}, K., {Hou}, Z., {Knox}, L., {et~al.}
\newblock {A Comparison of Cosmological Parameters Determined from CMB
  Temperature Power Spectra from the South Pole Telescope and the Planck
  Satellite}. 2017, \apj, 850, 101, \dodoi{10.3847/1538-4357/aa947b}

\bibitem[{{Addison} {et~al.}(2016){Addison}, {Huang}, {Watts}, {Bennett},
  {Halpern}, {Hinshaw}, \& {Weiland}}]{addison16}
{Addison}, G.~E., {Huang}, Y., {Watts}, D.~J., {Bennett}, C.~L., {Halpern}, M.,
  {Hinshaw}, G., \& {Weiland}, J.~L.
\newblock {Quantifying Discordance in the 2015 Planck CMB Spectrum}. 2016,
  \apj, 818, 132, \dodoi{10.3847/0004-637X/818/2/132}

\bibitem[{{Planck Collaboration} {et~al.}(2020{\natexlab{b}}){Planck
  Collaboration}, {Aghanim}, {Akrami}, {Ashdown}, {Aumont}, {Baccigalupi},
  {Ballardini}, {Banday}, {Barreiro}, {Bartolo}, {Basak}, {Battye}, {Benabed},
  {Bernard}, {Bersanelli}, {Bielewicz}, {Bock}, {Bond}, {Borrill}, {Bouchet},
  {Boulanger}, {Bucher}, {Burigana}, {Butler}, {Calabrese}, {Cardoso},
  {Carron}, {Challinor}, {Chiang}, {Chluba}, {Colombo}, {Combet}, {Contreras},
  {Crill}, {Cuttaia}, {de Bernardis}, {de Zotti}, {Delabrouille}, {Delouis},
  {Di Valentino}, {Diego}, {Dor{\'e}}, {Douspis}, {Ducout}, {Dupac}, {Dusini},
  {Efstathiou}, {Elsner}, {En{\ss}lin}, {Eriksen}, {Fantaye}, {Farhang},
  {Fergusson}, {Fernandez-Cobos}, {Finelli}, {Forastieri}, {Frailis},
  {Fraisse}, {Franceschi}, {Frolov}, {Galeotta}, {Galli}, {Ganga},
  {G{\'e}nova-Santos}, {Gerbino}, {Ghosh}, {Gonz{\'a}lez-Nuevo}, {G{\'o}rski},
  {Gratton}, {Gruppuso}, {Gudmundsson}, {Hamann}, {Handley}, {Hansen},
  {Herranz}, {Hildebrandt}, {Hivon}, {Huang}, {Jaffe}, {Jones}, {Karakci},
  {Keih{\"a}nen}, {Keskitalo}, {Kiiveri}, {Kim}, {Kisner}, {Knox},
  {Krachmalnicoff}, {Kunz}, {Kurki-Suonio}, {Lagache}, {Lamarre}, {Lasenby},
  {Lattanzi}, {Lawrence}, {Le Jeune}, {Lemos}, {Lesgourgues}, {Levrier},
  {Lewis}, {Liguori}, {Lilje}, {Lilley}, {Lindholm}, {L{\'o}pez-Caniego},
  {Lubin}, {Ma}, {Mac{\'\i}as-P{\'e}rez}, {Maggio}, {Maino}, {Mandolesi},
  {Mangilli}, {Marcos-Caballero}, {Maris}, {Martin}, {Martinelli},
  {Mart{\'\i}nez-Gonz{\'a}lez}, {Matarrese}, {Mauri}, {McEwen}, {Meinhold},
  {Melchiorri}, {Mennella}, {Migliaccio}, {Millea}, {Mitra},
  {Miville-Desch{\^e}nes}, {Molinari}, {Montier}, {Morgante}, {Moss}, {Natoli},
  {N{\o}rgaard-Nielsen}, {Pagano}, {Paoletti}, {Partridge}, {Patanchon},
  {Peiris}, {Perrotta}, {Pettorino}, {Piacentini}, {Polastri}, {Polenta},
  {Puget}, {Rachen}, {Reinecke}, {Remazeilles}, {Renzi}, {Rocha}, {Rosset},
  {Roudier}, {Rubi{\~n}o-Mart{\'\i}n}, {Ruiz-Granados}, {Salvati}, {Sandri},
  {Savelainen}, {Scott}, {Shellard}, {Sirignano}, {Sirri}, {Spencer},
  {Sunyaev}, {Suur-Uski}, {Tauber}, {Tavagnacco}, {Tenti}, {Toffolatti},
  {Tomasi}, {Trombetti}, {Valenziano}, {Valiviita}, {Van Tent}, {Vibert},
  {Vielva}, {Villa}, {Vittorio}, {Wand elt}, {Wehus}, {White}, {White},
  {Zacchei}, \& {Zonca}}]{planck18-6}
{Planck Collaboration}, {Aghanim}, N., {Akrami}, Y., {et~al.}
\newblock {Planck 2018 results. VI. Cosmological parameters}.
  2020{\natexlab{b}}, \aap, 641, A6, \dodoi{10.1051/0004-6361/201833910}

\bibitem[{{Riess} {et~al.}(2021){Riess}, {Casertano}, {Yuan}, {Bowers},
  {Macri}, {Zinn}, \& {Scolnic}}]{riess20}
{Riess}, A.~G., {Casertano}, S., {Yuan}, W., {Bowers}, J.~B., {Macri}, L.,
  {Zinn}, J.~C., \& {Scolnic}, D.
\newblock {Cosmic Distances Calibrated to 1\% Precision with Gaia EDR3
  Parallaxes and Hubble Space Telescope Photometry of 75 Milky Way Cepheids
  Confirm Tension with {\ensuremath{\Lambda}}CDM}. 2021, \apjl, 908, L6,
  \dodoi{10.3847/2041-8213/abdbaf}

\bibitem[{{Knox} \& {Millea}(2020)}]{knox19}
{Knox}, L., \& {Millea}, M.
\newblock {Hubble constant hunter's guide}. 2020, \prd, 101, 043533,
  \dodoi{10.1103/PhysRevD.101.043533}

\bibitem[{{Abazajian} {et~al.}(2016){Abazajian}, {Adshead}, {Ahmed}, {Allen},
  {Alonso}, {Arnold}, {Baccigalupi}, {Bartlett}, {Battaglia}, {Benson},
  {Bischoff}, {Borrill}, {Buza}, {Calabrese}, {Caldwell}, {Carlstrom}, {Chang},
  {Crawford}, {Cyr-Racine}, {De Bernardis}, {de Haan}, {di Serego Alighieri},
  {Dunkley}, {Dvorkin}, {Errard}, {Fabbian}, {Feeney}, {Ferraro}, {Filippini},
  {Flauger}, {Fuller}, {Gluscevic}, {Green}, {Grin}, {Grohs}, {Henning},
  {Hill}, {Hlozek}, {Holder}, {Holzapfel}, {Hu}, {Huffenberger}, {Keskitalo},
  {Knox}, {Kosowsky}, {Kovac}, {Kovetz}, {Kuo}, {Kusaka}, {Le Jeune}, {Lee},
  {Lilley}, {Loverde}, {Madhavacheril}, {Mantz}, {Marsh}, {McMahon},
  {Meerburg}, {Meyers}, {Miller}, {Munoz}, {Nguyen}, {Niemack}, {Peloso},
  {Peloton}, {Pogosian}, {Pryke}, {Raveri}, {Reichardt}, {Rocha}, {Rotti},
  {Schaan}, {Schmittfull}, {Scott}, {Sehgal}, {Shandera}, {Sherwin}, {Smith},
  {Sorbo}, {Starkman}, {Story}, {van Engelen}, {Vieira}, {Watson}, {Whitehorn},
  \& {Kimmy Wu}}]{abazajian16}
{Abazajian}, K.~N., {Adshead}, P., {Ahmed}, Z., {et~al.}
\newblock {CMB-S4 Science Book, First Edition}. 2016, ArXiv e-prints.
\newblock \doarXiv{1610.02743}

\bibitem[{{Simons Observatory Collaboration} {et~al.}(2019){Simons Observatory
  Collaboration}, {Ade}, {Aguirre}, {Ahmed}, {Aiola}, {Ali}, {Alonso},
  {Alvarez}, {Arnold}, {Ashton}, {Austermann}, {Awan}, {Baccigalupi},
  {Baildon}, {Barron}, {Battaglia}, {Battye}, {Baxter}, {Bazarko}, {Beall},
  {Bean}, {Beck}, {Beckman}, {Beringue}, {Bianchini}, {Boada}, {Boettger},
  {Bond}, {Borrill}, {Brown}, {Bruno}, {Bryan}, {Calabrese}, {Calafut},
  {Calisse}, {Carron}, {Challinor}, {Chesmore}, {Chinone}, {Chluba}, {Cho},
  {Choi}, {Coppi}, {Cothard}, {Coughlin}, {Crichton}, {Crowley}, {Crowley},
  {Cukierman}, {D'Ewart}, {D{\"u}nner}, {de Haan}, {Devlin}, {Dicker},
  {Didier}, {Dobbs}, {Dober}, {Duell}, {Duff}, {Duivenvoorden}, {Dunkley},
  {Dusatko}, {Errard}, {Fabbian}, {Feeney}, {Ferraro}, {Flux{\`a}}, {Freese},
  {Frisch}, {Frolov}, {Fuller}, {Fuzia}, {Galitzki}, {Gallardo}, {Tomas Galvez
  Ghersi}, {Gao}, {Gawiser}, {Gerbino}, {Gluscevic}, {Goeckner-Wald}, {Golec},
  {Gordon}, {Gralla}, {Green}, {Grigorian}, {Groh}, {Groppi}, {Guan},
  {Gudmundsson}, {Han}, {Hargrave}, {Hasegawa}, {Hasselfield}, {Hattori},
  {Haynes}, {Hazumi}, {He}, {Healy}, {Henderson}, {Hervias-Caimapo}, {Hill},
  {Hill}, {Hilton}, {Hilton}, {Hincks}, {Hinshaw}, {Hlo{\v{z}}ek}, {Ho}, {Ho},
  {Howe}, {Huang}, {Hubmayr}, {Huffenberger}, {Hughes}, {Ijjas}, {Ikape},
  {Irwin}, {Jaffe}, {Jain}, {Jeong}, {Kaneko}, {Karpel}, {Katayama}, {Keating},
  {Kernasovskiy}, {Keskitalo}, {Kisner}, {Kiuchi}, {Klein}, {Knowles},
  {Koopman}, {Kosowsky}, {Krachmalnicoff}, {Kuenstner}, {Kuo}, {Kusaka},
  {Lashner}, {Lee}, {Lee}, {Leon}, {Leung}, {Lewis}, {Li}, {Li}, {Limon},
  {Linder}, {Lopez-Caraballo}, {Louis}, {Lowry}, {Lungu}, {Madhavacheril},
  {Mak}, {Maldonado}, {Mani}, {Mates}, {Matsuda}, {Maurin}, {Mauskopf}, {May},
  {McCallum}, {McKenney}, {McMahon}, {Meerburg}, {Meyers}, {Miller},
  {Mirmelstein}, {Moodley}, {Munchmeyer}, {Munson}, {Naess}, {Nati},
  {Navaroli}, {Newburgh}, {Nguyen}, {Niemack}, {Nishino}, {Orlowski-Scherer},
  {Page}, {Partridge}, {Peloton}, {Perrotta}, {Piccirillo}, {Pisano},
  {Poletti}, {Puddu}, {Puglisi}, {Raum}, {Reichardt}, {Remazeilles},
  {Rephaeli}, {Riechers}, {Rojas}, {Roy}, {Sadeh}, {Sakurai}, {Salatino},
  {Sathyanarayana Rao}, {Schaan}, {Schmittfull}, {Sehgal}, {Seibert}, {Seljak},
  {Sherwin}, {Shimon}, {Sierra}, {Sievers}, {Sikhosana}, {Silva-Feaver},
  {Simon}, {Sinclair}, {Siritanasak}, {Smith}, {Smith}, {Spergel}, {Staggs},
  {Stein}, {Stevens}, {Stompor}, {Suzuki}, {Tajima}, {Takakura}, {Teply},
  {Thomas}, {Thorne}, {Thornton}, {Trac}, {Tsai}, {Tucker}, {Ullom},
  {Vagnozzi}, {van Engelen}, {Van Lanen}, {Van Winkle}, {Vavagiakis},
  {Verg{\`e}s}, {Vissers}, {Wagoner}, {Walker}, {Ward}, {Westbrook},
  {Whitehorn}, {Williams}, {Williams}, {Wollack}, {Xu}, {Yu}, {Yu}, {Zago},
  {Zhang}, {Zhu}, \& {The Simons Observatory collaboration}}]{ade19}
{Simons Observatory Collaboration}, {Ade}, P., {Aguirre}, J., {et~al.}
\newblock {The Simons Observatory: science goals and forecasts}. 2019, \jcap,
  2019, 056, \dodoi{10.1088/1475-7516/2019/02/056}

\bibitem[{{Galli} {et~al.}(2014){Galli}, {Benabed}, {Bouchet}, {Cardoso},
  {Elsner}, {Hivon}, {Mangilli}, {Prunet}, \& {Wandelt}}]{galli14}
{Galli}, S., {Benabed}, K., {Bouchet}, F., {Cardoso}, J.-F., {Elsner}, F.,
  {Hivon}, E., {Mangilli}, A., {Prunet}, S., \& {Wandelt}, B.
\newblock {CMB polarization can constrain cosmology better than CMB
  temperature}. 2014, \prd, 90, 063504, \dodoi{10.1103/PhysRevD.90.063504}

\bibitem[{{Trombetti} {et~al.}(2018){Trombetti}, {Burigana}, {De Zotti},
  {Galluzzi}, \& {Massardi}}]{trombetti18}
{Trombetti}, T., {Burigana}, C., {De Zotti}, G., {Galluzzi}, V., \& {Massardi},
  M.
\newblock {Average fractional polarization of extragalactic sources at Planck
  frequencies}. 2018, \aap, 618, A29, \dodoi{10.1051/0004-6361/201732342}

\bibitem[{{Gupta} {et~al.}(2019){Gupta}, {Reichardt}, {Ade}, {Anderson},
  {Archipley}, {Austermann}, {Avva}, {Beall}, {Bender}, {Benson}, {Bianchini},
  {Bleem}, {Carlstrom}, {Chang}, {Chiang}, {Citron}, {Moran}, {Crawford},
  {Crites}, {de Haan}, {Dobbs}, {Everett}, {Feng}, {Gallicchio}, {George},
  {Gilbert}, {Halverson}, {Harrington}, {Henning}, {Hilton}, {Holder},
  {Holzapfel}, {Hou}, {Hrubes}, {Huang}, {Hubmayr}, {Irwin}, {Knox}, {Lee},
  {Li}, {Lowitz}, {Luong-Van}, {Marrone}, {McMahon}, {Meyer}, {Mocanu}, {Mohr},
  {Montgomery}, {Nadolski}, {Natoli}, {Nibarger}, {Noble}, {Novosad}, {Padin},
  {Patil}, {Pryke}, {Ruhl}, {Saliwanchik}, {Sayre}, {Schaffer}, {Shirokoff},
  {Sievers}, {Smecher}, {Staniszewski}, {Stark}, {Story}, {Switzer}, {Tucker},
  {Vanderlinde}, {Veach}, {Vieira}, {Wang}, {Whitehorn}, {Williamson}, {Wu},
  {Yefremenko}, \& {Zhang}}]{gupta19}
{Gupta}, N., {Reichardt}, C.~L., {Ade}, P.~A.~R., {et~al.}
\newblock {Fractional polarization of extragalactic sources in the 500
  deg$^{2}$ SPTpol survey}. 2019, \mnras, 490, 5712,
  \dodoi{10.1093/mnras/stz2905}

\bibitem[{{Datta} {et~al.}(2019){Datta}, {Aiola}, {Choi}, {Devlin}, {Dunkley},
  {D{\"u}nner}, {Gallardo}, {Gralla}, {Halpern}, {Hasselfield}, {Hilton},
  {Hincks}, {Ho}, {Hubmayr}, {Huffenberger}, {Hughes}, {Kosowsky},
  {L{\'o}pez-Caraballo}, {Louis}, {Lungu}, {Marriage}, {Maurin}, {McMahon},
  {Moodley}, {Naess}, {Nati}, {Niemack}, {Page}, {Partridge}, {Prince},
  {Staggs}, {Switzer}, {Wollack}, \& {Farren}}]{datta19}
{Datta}, R., {Aiola}, S., {Choi}, S.~K., {et~al.}
\newblock {The Atacama Cosmology Telescope: two-season ACTPol extragalactic
  point sources and their polarization properties}. 2019, \mnras, 486, 5239,
  \dodoi{10.1093/mnras/sty2934}

\bibitem[{{Adachi} {et~al.}(2020){Adachi}, {Aguilar Fa{\'u}ndez}, {Arnold},
  {Baccigalupi}, {Barron}, {Beck}, {Bianchini}, {Chapman}, {Cheung}, {Chinone},
  {Crowley}, {Dobbs}, {El Bouhargani}, {Elleflot}, {Errard}, {Fabbian}, {Feng},
  {Fujino}, {Galitzki}, {Goeckner-Wald}, {Groh}, {Hall}, {Hasegawa}, {Hazumi},
  {Hirose}, {Jaffe}, {Jeong}, {Kaneko}, {Katayama}, {Keating}, {Kikuchi},
  {Kisner}, {Kusaka}, {Lee}, {Leon}, {Linder}, {Lowry}, {Matsuda}, {Matsumura},
  {Minami}, {Navaroli}, {Nishino}, {Pham}, {Poletti}, {Reichardt}, {Segawa},
  {Siritanasak}, {Tajima}, {Takakura}, {Takatori}, {Tanabe}, {Teply}, {Tsai},
  {Verg{\`e}s}, {Westbrook}, \& {Zhou}}]{polarbear20}
{Adachi}, S., {Aguilar Fa{\'u}ndez}, M.~A.~O., {Arnold}, K., {et~al.}
\newblock {A measurement of the CMB E-mode angular power spectrum at subdegree
  scales from 670 square degrees of POLARBEAR data}. 2020, arXiv e-prints,
  arXiv:2005.06168.
\newblock \doarXiv{2005.06168}

\bibitem[{{Dutcher} {et~al.}(2021){Dutcher}, {Balkenhol}, {Ade}, {Ahmed},
  {Anderes}, {Anderson}, {Archipley}, {Avva}, {Aylor}, {Barry}, {Basu Thakur},
  {Benabed}, {Bender}, {Benson}, {Bianchini}, {Bleem}, {Bouchet}, {Bryant},
  {Byrum}, {Carlstrom}, {Carter}, {Cecil}, {Chang}, {Chaubal}, {Chen}, {Cho},
  {Chou}, {Cliche}, {Crawford}, {Cukierman}, {Daley}, {de Haan}, {Denison},
  {Dibert}, {Ding}, {Dobbs}, {Everett}, {Feng}, {Ferguson}, {Foster}, {Fu},
  {Galli}, {Gambrel}, {Gardner}, {Goeckner-Wald}, {Gualtieri}, {Guns}, {Gupta},
  {Guyser}, {Halverson}, {Harke-Hosemann}, {Harrington}, {Henning}, {Hilton},
  {Hivon}, {Holder}, {Holzapfel}, {Hood}, {Howe}, {Huang}, {Irwin}, {Jeong},
  {Jonas}, {Jones}, {Khaire}, {Knox}, {Kofman}, {Korman}, {Kubik}, {Kuhlmann},
  {Kuo}, {Lee}, {Leitch}, {Lowitz}, {Lu}, {Meyer}, {Michalik}, {Millea},
  {Montgomery}, {Nadolski}, {Natoli}, {Nguyen}, {Noble}, {Novosad}, {Omori},
  {Padin}, {Pan}, {Paschos}, {Pearson}, {Posada}, {Prabhu}, {Quan},
  {Raghunathan}, {Rahlin}, {Reichardt}, {Riebel}, {Riedel}, {Rouble}, {Ruhl},
  {Sayre}, {Schiappucci}, {Shirokoff}, {Smecher}, {Sobrin}, {Stark}, {Stephen},
  {Story}, {Suzuki}, {Thompson}, {Thorne}, {Tucker}, {Umilta}, {Vale},
  {Vanderlinde}, {Vieira}, {Wang}, {Whitehorn}, {Wu}, {Yefremenko}, {Yoon}, \&
  {Young}}]{dutcher21}
{Dutcher}, D., {Balkenhol}, L., {Ade}, P.~A.~R., {et~al.}
\newblock {Measurements of the E-Mode Polarization and Temperature-E-Mode
  Correlation of the CMB from SPT-3G 2018 Data}. 2021, arXiv e-prints,
  arXiv:2101.01684.
\newblock \doarXiv{2101.01684}

\bibitem[{{Freedman} {et~al.}(2019){Freedman}, {Madore}, {Hatt}, {Hoyt},
  {Jang}, {Beaton}, {Burns}, {Lee}, {Monson}, {Neeley}, {Phillips}, {Rich}, \&
  {Seibert}}]{freedman19}
{Freedman}, W.~L., {Madore}, B.~F., {Hatt}, D., {Hoyt}, T.~J., {Jang}, I.-S.,
  {Beaton}, R.~L., {Burns}, C.~R., {Lee}, M.~G., {Monson}, A.~J., {Neeley},
  J.~R., {Phillips}, M.~M., {Rich}, J.~A., \& {Seibert}, M.
\newblock {The Carnegie-Chicago Hubble Program. VIII. An Independent
  Determination of the Hubble Constant Based on the Tip of the Red Giant
  Branch}. 2019, arXiv e-prints, arXiv:1907.05922.
\newblock \doarXiv{1907.05922}

\bibitem[{{Wong} {et~al.}(2020){Wong}, {Suyu}, {Chen}, {Rusu}, {Millon},
  {Sluse}, {Bonvin}, {Fassnacht}, {Taubenberger}, {Auger}, {Birrer}, {Chan},
  {Courbin}, {Hilbert}, {Tihhonova}, {Treu}, {Agnello}, {Ding}, {Jee},
  {Komatsu}, {Shajib}, {Sonnenfeld}, {Bland ford}, {Koopmans}, {Marshall}, \&
  {Meylan}}]{wong19}
{Wong}, K.~C., {Suyu}, S.~H., {Chen}, G. C.~F., {et~al.}
\newblock {H0LiCOW XIII. A 2.4\% measurement of H$_{0}$ from lensed quasars:
  5.3{\ensuremath{\sigma}} tension between early and late-Universe probes}.
  2020, \mnras, \dodoi{10.1093/mnras/stz3094}

\bibitem[{{Birrer} {et~al.}(2020){Birrer}, {Shajib}, {Galan}, {Millon}, {Treu},
  {Agnello}, {Auger}, {Chen}, {Christensen}, {Collett}, {Courbin}, {Fassnacht},
  {Koopmans}, {Marshall}, {Park}, {Rusu}, {Sluse}, {Spiniello}, {Suyu},
  {Wagner-Carena}, {Wong}, {Barnab{\`e}}, {Bolton}, {Czoske}, {Ding},
  {Frieman}, \& {Van de Vyvere}}]{birrer20}
{Birrer}, S., {Shajib}, A.~J., {Galan}, A., {et~al.}
\newblock {TDCOSMO IV: Hierarchical time-delay cosmography -- joint inference
  of the Hubble constant and galaxy density profiles}. 2020, arXiv e-prints,
  arXiv:2007.02941.
\newblock \doarXiv{2007.02941}

\bibitem[{{Jeong} {et~al.}(2014){Jeong}, {Chluba}, {Dai}, {Kamionkowski}, \&
  {Wang}}]{jeong14}
{Jeong}, D., {Chluba}, J., {Dai}, L., {Kamionkowski}, M., \& {Wang}, X.
\newblock {Effect of aberration on partial-sky measurements of the cosmic
  microwave background temperature power spectrum}. 2014, \prd, 89, 023003,
  \dodoi{10.1103/PhysRevD.89.023003}

\bibitem[{{Manzotti} {et~al.}(2014){Manzotti}, {Hu}, \&
  {Benoit-L{\'e}vy}}]{manzotti14}
{Manzotti}, A., {Hu}, W., \& {Benoit-L{\'e}vy}, A.
\newblock {Super-sample CMB lensing}. 2014, \prd, 90, 023003,
  \dodoi{10.1103/PhysRevD.90.023003}

\bibitem[{{Story} {et~al.}(2013){Story}, {Reichardt}, {Hou}, {Keisler}, {Aird},
  {Benson}, {Bleem}, {Carlstrom}, {Chang}, {Cho}, {Crawford}, {Crites}, {de
  Haan}, {Dobbs}, {Dudley}, {Follin}, {George}, {Halverson}, {Holder},
  {Holzapfel}, {Hoover}, {Hrubes}, {Joy}, {Knox}, {Lee}, {Leitch}, {Lueker},
  {Luong-Van}, {McMahon}, {Mehl}, {Meyer}, {Millea}, {Mohr}, {Montroy},
  {Padin}, {Plagge}, {Pryke}, {Ruhl}, {Sayre}, {Schaffer}, {Shaw}, {Shirokoff},
  {Spieler}, {Staniszewski}, {Stark}, {van Engelen}, {Vanderlinde}, {Vieira},
  {Williamson}, \& {Zahn}}]{story13}
{Story}, K.~T., {Reichardt}, C.~L., {Hou}, Z., {et~al.}
\newblock {A Measurement of the Cosmic Microwave Background Damping Tail from
  the 2500-Square-Degree SPT-SZ Survey}. 2013, \apj, 779, 86,
  \dodoi{10.1088/0004-637X/779/1/86}

\bibitem[{{Efstathiou} \& {Bond}(1999)}]{efstathiou99}
{Efstathiou}, G., \& {Bond}, J.~R.
\newblock Cosmic confusion: degeneracies among cosmological parameters derived
  from measurements of microwave background anisotropies. 1999, \mnras, 304,
  75.
\newblock
  \url{http://adsabs.harvard.edu/cgi-bin/nph-bib_query?bibcode=1999MNRAS.304...75E&db_key=AST}

\bibitem[{{Stompor} \& {Efstathiou}(1999)}]{stompor99}
{Stompor}, R., \& {Efstathiou}, G.
\newblock {Gravitational lensing of cosmic microwave background anisotropies
  and cosmological parameter estimation}. 1999, \mnras, 302, 735,
  \dodoi{10.1046/j.1365-8711.1999.02174.x}

\bibitem[{{Beutler} {et~al.}(2011){Beutler}, {Blake}, {Colless}, {Jones},
  {Staveley-Smith}, {Campbell}, {Parker}, {Saunders}, \& {Watson}}]{beutler11}
{Beutler}, F., {Blake}, C., {Colless}, M., {Jones}, D.~H., {Staveley-Smith},
  L., {Campbell}, L., {Parker}, Q., {Saunders}, W., \& {Watson}, F.
\newblock {The 6dF Galaxy Survey: baryon acoustic oscillations and the local
  Hubble constant}. 2011, \mnras, 416, 3017,
  \dodoi{10.1111/j.1365-2966.2011.19250.x}

\bibitem[{{Ross} {et~al.}(2015){Ross}, {Samushia}, {Howlett}, {Percival},
  {Burden}, \& {Manera}}]{ross15}
{Ross}, A.~J., {Samushia}, L., {Howlett}, C., {Percival}, W.~J., {Burden}, A.,
  \& {Manera}, M.
\newblock {The clustering of the SDSS DR7 main Galaxy sample - I. A 4 per cent
  distance measure at z = 0.15}. 2015, \mnras, 449, 835,
  \dodoi{10.1093/mnras/stv154}

\bibitem[{{Alam} {et~al.}(2017){Alam}, {Ata}, {Bailey}, {Beutler}, {Bizyaev},
  {Blazek}, {Bolton}, {Brownstein}, {Burden}, {Chuang}, {Comparat}, {Cuesta},
  {Dawson}, {Eisenstein}, {Escoffier}, {Gil-Mar{\'{\i}}n}, {Grieb}, {Hand},
  {Ho}, {Kinemuchi}, {Kirkby}, {Kitaura}, {Malanushenko}, {Malanushenko},
  {Maraston}, {McBride}, {Nichol}, {Olmstead}, {Oravetz}, {Padmanabhan},
  {Palanque-Delabrouille}, {Pan}, {Pellejero-Ibanez}, {Percival}, {Petitjean},
  {Prada}, {Price-Whelan}, {Reid}, {Rodr{\'{\i}}guez-Torres}, {Roe}, {Ross},
  {Ross}, {Rossi}, {Rubi{\~n}o-Mart{\'{\i}}n}, {Saito}, {Salazar-Albornoz},
  {Samushia}, {S{\'a}nchez}, {Satpathy}, {Schlegel}, {Schneider},
  {Sc{\'o}ccola}, {Seo}, {Sheldon}, {Simmons}, {Slosar}, {Strauss}, {Swanson},
  {Thomas}, {Tinker}, {Tojeiro}, {Maga{\~n}a}, {Vazquez}, {Verde}, {Wake},
  {Wang}, {Weinberg}, {White}, {Wood-Vasey}, {Y{\`e}che}, {Zehavi}, {Zhai}, \&
  {Zhao}}]{alam17}
{Alam}, S., {Ata}, M., {Bailey}, S., {et~al.}
\newblock {The clustering of galaxies in the completed SDSS-III Baryon
  Oscillation Spectroscopic Survey: cosmological analysis of the DR12 galaxy
  sample}. 2017, \mnras, 470, 2617, \dodoi{10.1093/mnras/stx721}

\bibitem[{{Blomqvist} {et~al.}(2019){Blomqvist}, {du Mas des Bourboux},
  {Busca}, {de Sainte Agathe}, {Rich}, {Balland}, {Bautista}, {Dawson},
  {Font-Ribera}, {Guy}, {Le Goff}, {Palanque-Delabrouille}, {Percival},
  {P{\'e}rez-R{\`a}fols}, {Pieri}, {Schneider}, {Slosar}, \&
  {Y{\`e}che}}]{blomqvist19}
{Blomqvist}, M., {du Mas des Bourboux}, H., {Busca}, N.~G., {et~al.}
\newblock {Baryon acoustic oscillations from the cross-correlation of
  Ly{\ensuremath{\alpha}} absorption and quasars in eBOSS DR14}. 2019, \aap,
  629, A86, \dodoi{10.1051/0004-6361/201935641}

\bibitem[{{Lewis} \& {Bridle}(2002)}]{lewis02b}
{Lewis}, A., \& {Bridle}, S.
\newblock {Cosmological parameters from CMB and other data: A Monte Carlo
  approach}. 2002, \prd, 66, 103511

\bibitem[{{Lewis} {et~al.}(2000){Lewis}, {Challinor}, \& {Lasenby}}]{lewis00}
{Lewis}, A., {Challinor}, A., \& {Lasenby}, A.
\newblock {Efficient Computation of Cosmic Microwave Background Anisotropies in
  Closed Friedmann-Robertson-Walker Models}. 2000, \apj, 538, 473,
  \dodoi{10.1086/309179}

\bibitem[{{Heymans} {et~al.}(2020){Heymans}, {Tr{\"o}ster}, {Asgari}, {Blake},
  {Hildebrandt}, {Joachimi}, {Kuijken}, {Lin}, {S{\'a}nchez}, {van den Busch},
  {Wright}, {Amon}, {Bilicki}, {de Jong}, {Crocce}, {Dvornik}, {Erben},
  {Getman}, {Giblin}, {Glazebrook}, {Hoekstra}, {Joudaki}, {Kannawadi},
  {Lidman}, {K{\"o}hlinger}, {Miller}, {Napolitano}, {Parkinson}, {Schneider},
  {Shan}, \& {Wolf}}]{heymans20}
{Heymans}, C., {Tr{\"o}ster}, T., {Asgari}, M., {et~al.}
\newblock {KiDS-1000 Cosmology: Multi-probe weak gravitational lensing and
  spectroscopic galaxy clustering constraints}. 2020, arXiv e-prints,
  arXiv:2007.15632.
\newblock \doarXiv{2007.15632}

\bibitem[{{Froustey} {et~al.}(2020){Froustey}, {Pitrou}, \&
  {Volpe}}]{froustey20}
{Froustey}, J., {Pitrou}, C., \& {Volpe}, M.~C.
\newblock {Neutrino decoupling including flavour oscillations and primordial
  nucleosynthesis}. 2020, \jcap, 2020, 015,
  \dodoi{10.1088/1475-7516/2020/12/015}

\bibitem[{{Bennett} {et~al.}(2020){Bennett}, {Buldgen}, {de Salas}, {Drewes},
  {Gariazzo}, {Pastor}, \& {Wong}}]{bennett20}
{Bennett}, J.~J., {Buldgen}, G., {de Salas}, P.~F., {Drewes}, M., {Gariazzo},
  S., {Pastor}, S., \& {Wong}, Y. Y.~Y.
\newblock {Towards a precision calculation of $N_{\rm eff}$ in the Standard
  Model II: Neutrino decoupling in the presence of flavour oscillations and
  finite-temperature QED}. 2020, arXiv e-prints, arXiv:2012.02726.
\newblock \doarXiv{2012.02726}

\bibitem[{{Abazajian} {et~al.}(2015){Abazajian}, {Arnold}, {Austermann},
  {Benson}, {Bischoff}, {Bock}, {Bond}, {Borrill}, {Calabrese}, {Carlstrom},
  {Carvalho}, {Chang}, {Chiang}, {Church}, {Cooray}, {Crawford}, {Dawson},
  {Das}, {Devlin}, {Dobbs}, {Dodelson}, {Dor{\'e}}, {Dunkley}, {Errard},
  {Fraisse}, {Gallicchio}, {Halverson}, {Hanany}, {Hildebrandt}, {Hincks},
  {Hlozek}, {Holder}, {Holzapfel}, {Honscheid}, {Hu}, {Hubmayr}, {Irwin},
  {Jones}, {Kamionkowski}, {Keating}, {Keisler}, {Knox}, {Komatsu}, {Kovac},
  {Kuo}, {Lawrence}, {Lee}, {Leitch}, {Linder}, {Lubin}, {McMahon}, {Miller},
  {Newburgh}, {Niemack}, {Nguyen}, {Nguyen}, {Page}, {Pryke}, {Reichardt},
  {Ruhl}, {Sehgal}, {Seljak}, {Sievers}, {Silverstein}, {Slosar}, {Smith},
  {Spergel}, {Staggs}, {Stark}, {Stompor}, {Vieregg}, {Wang}, {Watson},
  {Wollack}, {Wu}, {Yoon}, \& {Zahn}}]{snowmass13neutrinos}
{Abazajian}, K.~N., {Arnold}, K., {Austermann}, J., {et~al.}
\newblock {Neutrino physics from the cosmic microwave background and large
  scale structure}. 2015, Astroparticle Physics, 63, 66,
  \dodoi{10.1016/j.astropartphys.2014.05.014}

\bibitem[{Group {et~al.}(2020)Group, Zyla, Barnett, Beringer, Dahl, Dwyer,
  Groom, Lin, Lugovsky, Pianori, Robinson, Wohl, Yao, Agashe, Aielli, Allanach,
  Amsler, Antonelli, Aschenauer, Asner, Baer, Banerjee, Baudis, Bauer, Beatty,
  Belousov, Bethke, Bettini, Biebel, Black, Blucher, Buchmuller, Burkert,
  Bychkov, Cahn, Carena, Ceccucci, Cerri, Chakraborty, Chivukula, Cowan,
  D'Ambrosio, Damour, de~Florian, de~Gouvêa, DeGrand, de~Jong, Dissertori,
  Dobrescu, D'Onofrio, Doser, Drees, Dreiner, Eerola, Egede, Eidelman, Ellis,
  Erler, Ezhela, Fetscher, Fields, Foster, Freitas, Gallagher, Garren, Gerber,
  Gerbier, Gershon, Gershtein, Gherghetta, Godizov, Gonzalez-Garcia, Goodman,
  Grab, Gritsan, Grojean, Grünewald, Gurtu, Gutsche, Haber, Hanhart,
  Hashimoto, Hayato, Hebecker, Heinemeyer, Heltsley, Hernández-Rey, Hikasa,
  Hisano, Höcker, Holder, Holtkamp, Huston, Hyodo, Johnson, Kado, Karliner,
  Katz, Kenzie, Khoze, Klein, Klempt, Kowalewski, Krauss, Kreps, Krusche, Kwon,
  Lahav, Laiho, Lellouch, Lesgourgues, Liddle, Ligeti, Lippmann, Liss,
  Littenberg, Lourengo, Lugovsky, Lusiani, Makida, Maltoni, Mannel, Manohar,
  Marciano, Masoni, Matthews, Meißner, Mikhasenko, Miller, Milstead, Mitchell,
  Mönig, Molaro, Moortgat, Moskovic, Nakamura, Narain, Nason, Navas, Neubert,
  Nevski, Nir, Olive, Patrignani, Peacock, Petcov, Petrov, Pich, Piepke,
  Pomarol, Profumo, Quadt, Rabbertz, Rademacker, Raffelt, Ramani,
  Ramsey-Musolf, Ratcliff, Richardson, Ringwald, Roesler, Rolli, Romaniouk,
  Rosenberg, Rosner, Rybka, Ryskin, Ryutin, Sakai, Salam, Sarkar, Sauli,
  Schneider, Scholberg, Schwartz, Schwiening, Scott, Sharma, Sharpe, Shutt,
  Silari, Sjöstrand, Skands, Skwarnicki, Smoot, Soffer, Sozzi, Spanier,
  Spiering, Stahl, Stone, Sumino, Sumiyoshi, Syphers, Takahashi, Tanabashi,
  Tanaka, Taševský, Terashi, Terning, Thoma, Thorne, Tiator, Titov,
  Tkachenko, Tovey, Trabelsi, Urquijo, Valencia, Van~de Water, Varelas,
  Venanzoni, Verde, Vincter, Vogel, Vogelsang, Vogt, Vorobyev, Wakely,
  Walkowiak, Walter, Wands, Wascko, Weinberg, Weinberg, White, Wiencke,
  Willocq, Woody, Workman, Yokoyama, Yoshida, Zanderighi, Zeller, Zenin, Zhu,
  Zhu, Zimmermann, Anderson, Basaglia, Lugovsky, Schaffner, \& Zheng}]{pdg2020}
Group, P.~D., Zyla, P.~A., Barnett, R.~M., {et~al.}
\newblock {Review of Particle Physics}. 2020, Progress of Theoretical and
  Experimental Physics, 2020, \dodoi{10.1093/ptep/ptaa104}

\bibitem[{{Aver} {et~al.}(2020){Aver}, {Berg}, {Olive}, {Pogge}, {Salzer}, \&
  {Skillman}}]{aver20}
{Aver}, E., {Berg}, D.~A., {Olive}, K.~A., {Pogge}, R.~W., {Salzer}, J.~J., \&
  {Skillman}, E.~D.
\newblock {Improving Helium Abundance Determinations with Leo P as a Case
  Study}. 2020, arXiv e-prints, arXiv:2010.04180.
\newblock \doarXiv{2010.04180}

\bibitem[{{Dodelson} \& {Widrow}(1994)}]{dodelson94}
{Dodelson}, S., \& {Widrow}, L.~M.
\newblock {Sterile neutrinos as dark matter}. 1994, \prl, 72, 17,
  \dodoi{10.1103/PhysRevLett.72.17}

\bibitem[{{Lesgourgues} \& {Pastor}(2014)}]{lesgourgues14}
{Lesgourgues}, J., \& {Pastor}, S.
\newblock {Neutrino cosmology and Planck}. 2014, New Journal of Physics, 16,
  065002, \dodoi{10.1088/1367-2630/16/6/065002}

\bibitem[{Dolinski {et~al.}(2019)Dolinski, Poon, \& Rodejohann}]{dolinski19}
Dolinski, M.~J., Poon, A. W.~P., \& Rodejohann, W.
\newblock {Neutrinoless Double-Beta Decay: Status and Prospects}. 2019, Ann.
  Rev. Nucl. Part. Sci., 69, 219, \dodoi{10.1146/annurev-nucl-101918-023407}

\bibitem[{{Acero} {et~al.}(2019){Acero}, {Adamson}, {Aliaga}, {Alion},
  {Allakhverdian}, {Altakarli}, {Anfimov}, {Antoshkin}, {Aurisano}, {Back},
  {Backhouse}, {Baird}, {Balashov}, {Baldi}, {Bambah}, {Bashar}, {Bays},
  {Bending}, {Bernstein}, {Bhatnagar}, {Bhuyan}, {Bian}, {Blackburn}, {Blair},
  {Booth}, {Bour}, {Bromberg}, {Buchanan}, {Butkevich}, {Calvez}, {Campbell},
  {Carroll}, {Catano-Mur}, {Cedeno}, {Childress}, {Choudhary}, {Chowdhury},
  {Coan}, {Colo}, {Cooper}, {Corwin}, {Cremonesi}, {Davies}, {Derwent}, {Ding},
  {Djurcic}, {Doyle}, {Dukes}, {Duyang}, {Edayath}, {Ehrlich}, {Elkins},
  {Feldman}, {Filip}, {Flanagan}, {Frank}, {Gallagher}, {Gandrajula}, {Gao},
  {Germani}, {Giri}, {Gomes}, {Goodman}, {Grichine}, {Groh}, {R. Group}, {Guo},
  {Habig}, {Hakl}, {Hartnell}, {Hatcher}, {Hatzikoutelis}, {Heller}, {Hewes},
  {Himmel}, {Holin}, {Howard}, {Huang}, {Hylen}, {Jediny}, {Johnson}, {Judah},
  {Kakorin}, {Kalra}, {Kaplan}, {Keloth}, {Klimov}, {Koerner}, {Kolupaeva},
  {Kotelnikov}, {Kreymer}, {Kulenberg}, {Kumar}, {Kuruppu}, {Kus}, {Lackey},
  {Lang}, {Lin}, {Lokajicek}, {Lozier}, {Luchuk}, {Maan}, {Magill}, {Mann},
  {Marshak}, {Martinez-Casales}, {Matveev}, {Mendez}, {Messier}, {Meyer},
  {Miao}, {Miller}, {Mishra}, {Mislivec}, {Mohanta}, {Moren}, {Mualem},
  {Muether}, {Mufson}, {Mulder}, {Murphy}, {Musser}, {Naples}, {Nayak},
  {Nelson}, {Nichol}, {Nikseresht}, {Niner}, {Norman}, {Nosek}, {Olshevskiy},
  {Olson}, {Paley}, {Patterson}, {Pawloski}, {Pershey}, {Petrova}, {Petti},
  {Phan}, {Plunkett}, {Potukuchi}, {Principato}, {Psihas}, {Radovic}, {Raj},
  {Rameika}, {Rebel}, {Rojas}, {Ryabov}, {Samoylov}, {Sanchez}, {Sanchez
  Falero}, {Seong}, {Shanahan}, {Sheshukov}, {Singh}, {Singh}, {Smith},
  {Smolik}, {Snopok}, {Solomey}, {Song}, {Sousa}, {Soustruznik}, {Strait},
  {Suter}, {Sutton}, {Talaga}, {Tapia Oregui}, {Tas}, {Thayyullathil},
  {Thomas}, {Tiras}, {Torbunov}, {Tripathi}, {Tsaris}, {Torun}, {Urheim},
  {Vahle}, {Vasel}, {Vinton}, {Vokac}, {Vrba}, {Wallbank}, {Wang}, {Warburton},
  {Wetstein}, {While}, {Whittington}, {Wojcicki}, {Wolcott}, {Yadav}, {Yallappa
  Dombara}, {Yonehara}, {Yu}, {Zadorozhnyy}, {Zalesak}, {Zamorano}, \&
  {Zwaska}}]{acero19}
{Acero}, M.~A., {Adamson}, P., {Aliaga}, L., {et~al.}
\newblock {First measurement of neutrino oscillation parameters using neutrinos
  and antineutrinos by NOvA}. 2019, arXiv e-prints, arXiv:1906.04907.
\newblock \doarXiv{1906.04907}

\bibitem[{{Bianchini} {et~al.}(2020){Bianchini}, {Wu}, {Ade}, {Anderson},
  {Austermann}, {Avva}, {Beall}, {Bender}, {Benson}, {Bleem}, {Carlstrom},
  {Chang}, {Chaubal}, {Chiang}, {Citron}, {Moran}, {Crawford}, {Crites}, {de
  Haan}, {Dobbs}, {Everett}, {Gallicchio}, {George}, {Gilbert}, {Gupta},
  {Halverson}, {Harrington}, {Henning}, {Hilton}, {Holder}, {Holzapfel},
  {Hrubes}, {Huang}, {Hubmayr}, {Irwin}, {Knox}, {Lee}, {Li}, {Lowitz},
  {Manzotti}, {McMahon}, {Meyer}, {Millea}, {Mocanu}, {Montgomery}, {Nadolski},
  {Natoli}, {Nibarger}, {Noble}, {Novosad}, {Omori}, {Padin}, {Patil}, {Pryke},
  {Reichardt}, {Ruhl}, {Saliwanchik}, {Sayre}, {Schaffer}, {Sievers}, {Simard},
  {Smecher}, {Stark}, {Story}, {Tucker}, {Vanderlinde}, {Veach}, {Vieira},
  {Wang}, {Whitehorn}, \& {Yefremenko}}]{bianchini20}
{Bianchini}, F., {Wu}, W.~L.~K., {Ade}, P.~A.~R., {et~al.}
\newblock {Constraints on Cosmological Parameters from the 500 deg$^{2}$ SPTPOL
  Lensing Power Spectrum}. 2020, \apj, 888, 119,
  \dodoi{10.3847/1538-4357/ab6082}

\bibitem[{{Planck Collaboration} {et~al.}(2014){Planck Collaboration}, {Ade},
  {Aghanim}, {Armitage-Caplan}, {Arnaud}, {Ashdown}, {Atrio-Barandela},
  {Aumont}, {Baccigalupi}, {Banday}, \& et~al.}]{planck13-16}
{Planck Collaboration}, {Ade}, P.~A.~R., {Aghanim}, N., {Armitage-Caplan}, C.,
  {Arnaud}, M., {Ashdown}, M., {Atrio-Barandela}, F., {Aumont}, J.,
  {Baccigalupi}, C., {Banday}, A.~J., \& et~al.
\newblock {Planck 2013 results. XVI. Cosmological parameters}. 2014, \aap, 571,
  A16, \dodoi{10.1051/0004-6361/201321591}

\bibitem[{{Planck Collaboration} {et~al.}(2016{\natexlab{b}}){Planck
  Collaboration}, {Ade}, {Aghanim}, {Arnaud}, {Ashdown}, {Aumont},
  {Baccigalupi}, {Banday}, {Barreiro}, {Bartlett}, \& et~al.}]{planck15-13}
{Planck Collaboration}, {Ade}, P.~A.~R., {Aghanim}, N., {Arnaud}, M.,
  {Ashdown}, M., {Aumont}, J., {Baccigalupi}, C., {Banday}, A.~J., {Barreiro},
  R.~B., {Bartlett}, J.~G., \& et~al.
\newblock {Planck 2015 results. XIII. Cosmological parameters}.
  2016{\natexlab{b}}, \aap, 594, A13, \dodoi{10.1051/0004-6361/201525830}

\bibitem[{{Addison}(2021)}]{addison21}
{Addison}, G.~E.
\newblock {High $H_0$ Values from CMB E-mode Data: A Clue for Resolving the
  Hubble Tension?} 2021, arXiv e-prints, arXiv:2102.00028.
\newblock \doarXiv{2102.00028}

\bibitem[{Pordes {et~al.}(2007)}]{pordes07}
Pordes, R., {et~al.}
\newblock {The Open Science Grid}. 2007, J. Phys. Conf. Ser., 78, 012057,
  \dodoi{10.1088/1742-6596/78/1/012057}

\bibitem[{{Sfiligoi} {et~al.}(2009){Sfiligoi}, {Bradley}, {Holzman},
  {Mhashilkar}, {Padhi}, \& {Wurthwein}}]{sfiligoi09}
{Sfiligoi}, I., {Bradley}, D.~C., {Holzman}, B., {Mhashilkar}, P., {Padhi}, S.,
  \& {Wurthwein}, F.
\newblock The Pilot Way to Grid Resources Using glideinWMS. 2009, in 2, Vol.~2,
  2009 WRI World Congress on Computer Science and Information Engineering,
  428--432, \dodoi{10.1109/CSIE.2009.950}

\bibitem[{Hunter(2007)}]{hunter07}
Hunter, J.~D.
\newblock Matplotlib: A 2D graphics environment. 2007, Computing In Science \&
  Engineering, 9, 90, \dodoi{10.1109/MCSE.2007.55}

\bibitem[{Jones {et~al.}(2001)Jones, Oliphant, Peterson, {et~al.}}]{jones01}
Jones, E., Oliphant, T., Peterson, P., {et~al.} 2001, {SciPy}: Open source
  scientific tools for {Python}.
\newblock \url{http://www.scipy.org/}

\bibitem[{van~der Walt {et~al.}(2011)van~der Walt, Colbert, \&
  Varoquaux}]{vanDerWalt11}
van~der Walt, S., Colbert, S., \& Varoquaux, G.
\newblock The NumPy Array: A Structure for Efficient Numerical Computation.
  2011, Computing in Science Engineering, 13, 22, \dodoi{10.1109/MCSE.2011.37}

\bibitem[{{Motloch} \& {Hu}(2019)}]{motloch19}
{Motloch}, P., \& {Hu}, W.
\newblock {Lensing covariance on cut sky and SPT-Planck lensing tensions}.
  2019, \prd, 99, 023506, \dodoi{10.1103/PhysRevD.99.023506}

\end{thebibliography}

\end{document}